\documentclass[a4paper,11pt]{article}
\pdfoutput=1 

\usepackage{jheppub} 
\usepackage{amsmath,amssymb,amsthm,amscd,graphicx,mathtools}

\input epsf.sty

\makeatletter
\newsavebox{\@brx}
\newcommand{\llangle}[1][]{\savebox{\@brx}{\(\m@th{#1\langle}\)}%
  \mathopen{\copy\@brx\kern-0.5\wd\@brx\usebox{\@brx}}}
\newcommand{\rrangle}[1][]{\savebox{\@brx}{\(\m@th{#1\rangle}\)}%
  \mathclose{\copy\@brx\kern-0.5\wd\@brx\usebox{\@brx}}}
\makeatother

\theoremstyle{definition}

\newcommand{\re}{{\rm e}}
\newcommand{\ri}{{\rm i}}
\newcommand{\rd}{{\rm d}}
\newcommand{\mx}{{\mathsf{x}}}
\newcommand{\my}{{\mathsf{y}}}

\newcommand{\bxi}{{\boldsymbol \xi}}
\newcommand{\sfu}{{\mathsf{u}}}
\newcommand{\sfv}{{\mathsf{v}}}

\newcommand{\be}{\begin{equation}}
\newcommand{\ee}{\end{equation}}
\newcommand{\ba}{\begin{aligned}}
\newcommand{\ea}{\end{aligned}}
\newcommand{\ben}{\begin{eqnarray}\displaystyle}
\newcommand{\een}{\end{eqnarray}}

\newcommand{\sectiono}[1]{\section{#1}\setcounter{equation}{0}}

\newdimen\tableauside\tableauside=1.0ex
\newdimen\tableaurule\tableaurule=0.4pt
\newdimen\tableaustep
\def\phantomhrule#1{\hbox{\vbox to0pt{\hrule height\tableaurule width#1\vss}}}
\def\phantomvrule#1{\vbox{\hbox to0pt{\vrule width\tableaurule height#1\hss}}}
\def\sqr{\vbox{%
  \phantomhrule\tableaustep
  \hbox{\phantomvrule\tableaustep\kern\tableaustep\phantomvrule\tableaustep}%
  \hbox{\vbox{\phantomhrule\tableauside}\kern-\tableaurule}}}
\def\squares#1{\hbox{\count0=#1\noindent\loop\sqr
  \advance\count0 by-1 \ifnum\count0>0\repeat}}
\def\tableau#1{\vcenter{\offinterlineskip
  \tableaustep=\tableauside\advance\tableaustep by-\tableaurule
  \kern\normallineskip\hbox
    {\kern\normallineskip\vbox
      {\gettableau#1 0 }%
     \kern\normallineskip\kern\tableaurule}%
  \kern\normallineskip\kern\tableaurule}}
\def\gettableau#1{\ifnum#1=0\let\next=\null\else
\squares{#1}\let\next=\gettableau\fi\next}

\title{\huge{Quantized mirror curves and resummed WKB}}

\author{
Szabolcs Zakany
}

\affiliation{D\'epartement de Physique Th\'eorique\\
Universit\'e de Gen\`eve, Gen\`eve, CH-1211 Switzerland}

\emailAdd{szabolcs.zakany@unige.ch} 

\abstract{Based on previous insights, we present an ansatz to obtain quantization conditions and eigenfunctions for a family of difference equations which arise from quantized mirror curves in the context of local mirror symmetry of toric Calabi-Yau threefolds.
It is a first principles construction, which yields closed expressions for the quantization conditions and  the eigenfunctions when $\hbar/2\pi \in \mathbb Q$. The key ingredient is the modular duality structure of the underlying quantum integrable system. We use our ansatz to write down explicit results in some examples, which are successfully checked against purely numerical results for both the spectrum and the eigenfunctions. Concerning the quantization conditions, we also provide evidence that, in the rational case, this method yields a resummation of conjectured quantization conditions involving enumerative invariants of the underlying toric Calabi-Yau threefold.}

\begin{document}
\maketitle
\flushbottom

 
 \sectiono{Introduction}
 
 Finding eigenfunctions and eigenvalues of differential or difference operators is an ubiquitous problem in physics. Finding exact expressions for them is most of the time rather difficult, even in the one dimensional case where the operator acts on functions on the real line (the setup often considered in standard quantum mechanics). 
 The operators considered in this work are built as Laurent polynomials of the exponentials
 \be
  \sfu=\re^{\mx}, \qquad \sfv = \re^{\my},
  \ee
 where the operators $\mx$ and $\my$ satisfy the canonical commutation relation $[\mx,\my]=\ri \hbar$ for $\hbar \in \mathbb R$. The operators $\sfu$ and $\sfv$ can be represented as multiplication and shift operators acting on functions on $\mathbb R$. Polynomials of $\sfu$ and $\sfv$ and their inverses correspond to more general difference operators acting on functions.
 Such operators arise in several areas of theoretical physics. One example is the quantization of the spectral curves of some particular integrable systems, yielding the Baxter equation which is central in the study of those systems.  Another is the quantization of mirror curves.
 When the genus $g_\Sigma$ of the underlying curve is greater or equal to one, the operator may have a discrete spectrum. More precisely, it has been shown in many examples that the inverse operator of the quantized curve is a positive definite trace class  self-adjoint operator acting on $L^2(\mathbb R)$ \cite{KaMa,lst},
implying that the spectrum of the operator is discrete. 

Our main motivation for the study of such difference operators stems from local mirror symmetry for toric Calabi-Yau (CY) threefolds. It is well known that the mirror geometry of a toric CY threefold is encoded in a curve in ${\mathbb C}^* \times {\mathbb C}^* $ called the mirror curve. Period integrals on this curve give the K\"ahler parameters in term of the complex structure parameters, as well as the derivatives of the prepotential. 
More recently, it has been proposed that an appropriate quantization of this same mirror curve leads to interesting deformations of the prepotential, which are tightly related to topological strings and enumerative invariants of the toric CY threefold \cite{adkmv,acdkv}. Building on insights coming from ABJ(M) theory, this idea has been further developed, in particular in \cite{hmmo, kallenmarino}.
The most complete realization of this idea is the so called ``Topological String/Spectral Theory" (TS/ST) correspondence \cite{ghm, cgm} (for a review, see \cite{mReview}), whose extension to the open sector has been undertaken in \cite{mz1,mz2}. 

In the context of the TS/ST correspondence, the central object to consider is an operator acting on $L^2(\mathbb R)$, essentially given by the quantization of the mirror curve of the underlying CY threefold.  
The main idea is that spectral quantities of this operator can be written down explicitly using enumerative invariants of the underlying geometry\footnote{These enumerative invariants essentially boil down to BPS numbers, counting BPS states of an M-theoretic lift of the topological string.}.
The operator itself leads to a difference equation, which takes the form we consider here.
In fact, all examples of difference equations that we will consider come from geometries studied in this context. That is why, in the following, we will call each example by the name of the underlying toric CY threefold. The TS/ST correspondence is in a very large measure yet unproven, so it has to be explicitly tested case by case (see for example \cite{ghm, ghm2, cgm,gkmr,cGum,mzOLD,kmzOLD}). Such tests require the knowledge of the various spectral quantities of the operator, as for example the eigenvalues and eigenfunctions of the corresponding difference equation. 
It is therefore highly desirable to have an explicit and first principle way of computing these objects. 

In the original TS/ST approach, one considers a spectral problem in one dimension for the (inverse) quantized mirror curve operator. 
The quantization condition proposed in \cite{ghm,cgm} for the spectrum gives a codimension one manifold in the space of moduli of the curve. A proposal for the eigenfunctions was put forward in \cite{mz1,mz2}.
However, there is another spectral problem related to  the mirror curve, given by the associated Goncharov-Kenyon integrable system, or cluster integrable system \cite{GK}. 
It is an integrable system with $g_\Sigma$ mutually commuting hamiltonians, where $g_\Sigma$ is the genus of the mirror curve.
This defines a spectral problem in $g_\Sigma$ dimensions.
The quantization conditions giving the energies of the joint eigenstates of the integrable system were proposed in \cite{wzh,fhm}, and give a discrete subset of the codimension one manifold of \cite{cgm}.
In fact, the difference equation given by the quantized mirror curve (i.e. the spectral problem in one dimension) corresponds to the Baxter equation of the cluster integrable system. Based on the study of other integrable systems, as was done for example in the case of the non-relativistic Toda lattice in \cite{gutzwiller,gp,kl}, it is natural to expect that the appropriate solutions of the Baxter equation satisfy more restrictive boundary conditions when the moduli of the curve correspond to a joint eigenstate of the integrable system.
More precisely, we expect that for special values of the moduli, one of the square-integrable eigenfunctions of the quantized mirror curve operator decays more rapidly at plus or minus infinity, thus signalling a joint eigenstate of the corresponding integrable system. This phenomenon was checked in \cite{mz2} in the case of the mirror curve of the geometry known as the resolution of $\mathbb C^3/\mathbb Z_5$, for $\hbar=2\pi$.
This subset of solutions of the difference equation with enhanced decay will be called ``fully on-shell", meaning that values of the true moduli are ``on-shell" from the point of view of the integrable system.

The aim of this paper is to show how to construct these ``fully on-shell" eigenfunctions directly from the difference equation.
This can be explicitly done when $\hbar$ is of the form  $\hbar/2\pi \in \mathbb Q$. We will refer to this as the ``rational case".

The solution presented here is an ansatz which is based on several recent developments. It was noticed in \cite{wzh} that the quantization conditions given in that reference can be written in a way that is for the most part invariant under $\hbar/2\pi \leftrightarrow 2\pi/\hbar$. 
The importance of this duality was also emphasised in \cite{hatsuda1}. This duality is to be considered as a manifestation of the ``modular double" structure \cite{faddeev1} of the corresponding integrable system (see \cite{klsts} for the case of the relativistic Toda lattice). It is then natural to assume that the invariance under $\hbar/2\pi \leftrightarrow 2\pi/\hbar$ should not only be manifest in the spectrum of the difference equation, but also in its eigenfunctions, at least for the subset which is ``fully on-shell". This is the point of view taken in \cite{KP}, where an ansatz is proposed for the eigenfunctions based on this consideration. 
Concretely, the ansatz of \cite{KP} consists in taking the resummed WKB eigenfunction and symmetrize it with respect to $\hbar/2\pi \leftrightarrow 2\pi/\hbar$. Monodromy invariance of the eigenfunction thus constructed yields the quantization conditions of \cite{wzh}. Unfortunately, this idea alone does not allow us to get explicit, closed expressions for the ``fully on-shell" eigenfunction, because the resummed WKB expression is not known exactly, but only, for example, as a large $X=\re^x$ expansion.
We argue here, and show in many examples, that this resummation can be performed in the rational case, that is, when $\hbar/2\pi \in \mathbb Q$. Technically, the method to perform this resummation can be seen as a refinement of what is done in \cite{hsx,hkt} to obtain the relation between the spectrum at $\hbar$ and at its dual value $4\pi^2/\hbar$. In particular, a certain matrix appearing in those references plays an important role.
Our method is also inspired by \cite{ks}, even if in that reference the authors mainly consider the case where $\hbar$ has an imaginary part\footnote{The common domain of validity of our method and the results of \cite{ks} is the case $\hbar=2\pi$, where we find agreement between the two solutions. At the time of the writing of \cite{ks}, it has already been suggested to us by one of the authors, \mbox{R. Kashaev}, that the case $\hbar/2\pi \in \mathbb Q$ could be somehow addressed using a matrix-based method.}.

As already alluded to, given any toric CY threefold, one can consider the corresponding cluster integrable system. 
The archetypical example is the relativistc Toda lattice of $N$ particles, which is the cluster integrable system associated to the  toric CY threefolds often called the resolved $A_{N-1}$ geometries. For the relativistic Toda lattice, it has been shown in \cite{klsts} using the Quantum Inverse Scattering Method, that the joint eigenfunctions of the integrable system (the spectral problem in $g_\Sigma$ dimensions) can be explicitly built from the ``fully on-shell" eigenfunctions of the Baxter equation through a certain integral transform. Our method should thus provide a solution for the relativistic Toda lattice eigenfunctions for the rational case. 
Let us mention that the eigenfunctions for the relativistic Toda have been constructed in \cite{sciarappa1,sciarappa2}, using a different method relying on gauge theory computations of instanton partition functions in the presence of defects. That construction gives a solution for any value of $\hbar$, but only as an expansion in an auxiliary parameter.

The paper is structured as follows: in section \ref{sect2}, we present the family of difference equations, and study them using the WKB ansatz. In section \ref{sect3}, we derive the main result, and present the different formulas to compute its components.
Section \ref{sect4} is devoted to examples and tests of the formula. In section \ref{sect5}, we investigate the relationship between our quantization condition and the conjectural one of \cite{wzh, hm, fhm}. In the last section, we give some concluding remarks.

 
 \sectiono{About the difference equation}
 \label{sect2}
 
 We introduce the family of difference equations. 
 They are then studied using the WKB ansatz, yielding the WKB eigenfunction, which is central in our construction. In this section, we do not yet assume that $\hbar/2\pi$ is rational.
 For simplicity, we will focus on the cases where the difference equation is of second order, which we call ``hyperelliptic cases" because the underlying Riemann surface is a two-sheeted cover of the plane. 
 But the method is in principle also applicable to higher order cases.
 
 \subsection{The difference equation and its dual}
 
 Since our main motivation is the TS/ST correspondence, our starting point is a mirror curve of genus $g_\Sigma \geq 1$, defined by a curve
\be
	W(x,y)=0,
\ee
where $W(x,y)$ is a Laurent polynomial of $\re^x$ and $\re^y$ with $N$ monomials, and can take the form
\be
	W(x,y)=\sum_{k=1}^{N-1} \xi_{k} \re^{\mu_{k}x+\nu_{k} y}+\kappa.
\ee
The $\mu_n$ and $\nu_n$ are integer numbers. The parameter $\kappa$ is a true modulus of the curve, whereas the parameters $\xi_n$ can be mass parameters as well as other true moduli, depending on the curve\footnote{For a curve of genus $g_\Sigma$, there are $g_\Sigma$ ``true" moduli, and the remaining ones are usually called ``mass parameters". They are associated to non-compact cycles in the mirror geometry and lead to trivial mirror maps.}.
By performing constant shifts of $x$ and $y$ and by an overall scaling of the equation defining the mirror curve, we can always set three of the parameters $\xi_n$ to $1$.
These curves are typical in the study of mirror symmetry of toric Calabi-Yau threefolds.
We quantize the curve by promoting $x$ and $y$ to canonically commuting self-adjoint operators (and use Weyl prescription to fix ordering ambiguities):
\be
	x,y \rightarrow {\mx},{\my }, \qquad \qquad [{\mx},{\my }]=\ri \hbar, \qquad \qquad \hbar \in \mathbb R.
\ee 
This quantization procedure is the one used in the TS/ST correspondence. It gives the operator $W(\mx,\my)$, which can be expressed as a polynomial of $\re^{\mx}$ and $\re^{\my}$ using the BCH formula to split exponentials. In the $x$ representation, the operator $\my$ acts as a derivative $-\ri \partial_x$. The operator $\re^{\nu_k \my}$ acts as a shift operator on a function, since its action amounts to Taylor expanding it. 
The operator given by
\be
	\label{Odef}
	{\mathsf O}=\sum_{k=1}^{N-1} \xi_{k} \re^{\mu_{k}\mx+\nu_{k} \my}
\ee
is then a difference operator acting on functions. Its inverse
\be
	\rho={\mathsf O}^{-1}
\ee
has nice properties: it has been shown in many examples in \cite{KaMa, lst} that it is a positive definite trace class  self-adjoint operator on $L^2(\mathbb R)$, at least for appropriate values of $\xi_k$. As such, it admits a discrete set of eigenvalues $(-\kappa_n)^{-1}$ and eigenfunctions $\psi_n(x)$. An eigenfunction $\psi_n(x)$ is in the kernel of the operator $W(\mx,\my)$ given by states $|\psi \rangle$ such that
\be
	\label{opWequals0}
	W(\mx,\my) | \psi \rangle=0,
\ee
but not necessarily all functions in this kernel are eigenfunctions of $\rho$ since they may not be in the image of $\rho$. Indeed, the operator $\mathsf O$ should really be considered as the inverse of $\rho$, so its domain should be restricted to the image $\rho(L^2(\mathbb R))$. We will sometimes call the eigenfunctions $\psi_n(x)$ ``on-shell" eigenfunctions, and other functions $\psi(x)$ in the kernel of $W(\mx,\my)$ will be called ``off-shell" eigenfunctions by abuse of language (since they are not truly eigenfunctions of $\rho$). 
All these functions satisfy the difference equation
\be
	\label{diffeq1}
	\sum_{n=1}^{N-1}\xi_{k} \, \re^{-\mu_k \nu_k \frac{\ri \hbar  }{2}} \, \re^{\mu_{n}x} \psi(x-\nu_n \ri \hbar)+\kappa \psi(x)=0,
\ee
but only for the specific values $\kappa=\kappa_n$ do we find on-shell eigenfunctions.
We will focus on hyperelliptic curves, where the difference equation is of order 2.

In the case where the underlying mirror curve is of genus $g_\Sigma$ greater then 1, we have $g_\Sigma$ true moduli, $g_\Sigma-1$ of which are among the $\xi_k$. In principle, they could act as spectator parameters (like the other $\xi_k$) and take arbitrary values. As we will see in the examples, our method only gives the eigenfunctions for some specific values of these true moduli which, in this sense, are also quantized. Their values turn out to correspond to the joint eigenstates of the corresponding integrable system (the $g_\Sigma$ commuting hamiltonians), and the eigenfunctions thus obtained are the ``fully on-shell" eigenfunctions.
 
The spectrum and eigenfunctions of the operator $\rho$ can be numerically well approximated using the difference equation (\ref{diffeq1}) and a hamiltonian truncation method on a basis of $ L^2(\mathbb R)$, which we take to be the harmonic oscillator basis. All the numerical results used to test against  the proposed exact expressions are generated using this technique.
\newline

These difference equations have an interesting modular duality structure, as pointed out in the seminal paper \cite{faddeev1}.
As already implied, equation (\ref{opWequals0}) can be written as the eigenvalue equation
\be
	\mathsf O  | \psi \rangle = -\kappa | \psi \rangle,
\ee
with $\mathsf O$ given by (\ref{Odef}).
The point is that there exist two other operators which formally commute with this operator, and so all the algebra spanned by them commutes with $\mathsf O$.
Indeed, for any integers $m,n$, we formally have
\be
	[\re^{ m \mathsf x},\re^{n \frac{2\pi}{\hbar}\mathsf y}] = 0, \qquad \qquad [\re^{ m \mathsf y},\re^{n \frac{2\pi}{\hbar}\mathsf x}] = 0.
\ee
For example, the following operator $\tilde {\mathsf O}$, which we call the dual operator
\be
	\tilde {\mathsf O} = \sum_{n=1}^{N-1} \tilde \xi_{n} \re^{\mu_{n} \frac{2\pi}{\hbar}\mx+\nu_{n} \frac{2\pi}{\hbar} \my},
\ee
formally commutes with ${\mathsf O}$, and so we may expect that it can be diagonalized by some of the eigenfunctions of the operator $\mathsf O$.
We assume that this is true for the ``fully on-shell" eigenfunctions.
 The eigenvalue of $\tilde {\mathsf O}$ will be denoted $-\tilde \kappa$. The operator $\tilde {\mathsf O}$ has essentially the same form as  $ {\mathsf O}$, but a priori the different moduli and parameters can take arbitrary values. In the following, we will see some examples of relations between $\kappa$ and $\tilde \kappa$ (and $\xi_n$ and $\tilde \xi_n$), and how they arise concretely. 
The difference equation given by the dual operator is
\be
	\sum_{n=1}^{N-1} \tilde \xi_{n} \, \re^{-\mu_n \nu_n \frac{\ri \hbar  }{2} \frac{4\pi^2}{\hbar^2}} \, \re^{\mu_{n} \frac{2\pi x}{\hbar}} \psi(x-2\pi \ri \nu_n)+\tilde \kappa \psi(x)=0.
\ee
By rescaling the eigenfunction $\tilde \psi(x) = \psi \left (\frac{\hbar}{2\pi} x \right )$, and renaming
\be
	\label{dualvars}
	\hbar_D=\frac{4\pi^2}{\hbar}, \qquad \qquad x_D=\frac{2\pi x}{\hbar},
\ee
this gives
\be
	\label{dualDiffEq1}
	\sum_{n=1}^{N-1} \tilde \xi_{n} \, \re^{-\mu_n \nu_n \frac{\ri \hbar_D  }{2}} \, \re^{\mu_{n}x_D} \tilde \psi(x_D- \nu_n \ri \hbar_D)+\tilde \kappa \tilde \psi(x_D)=0,
\ee
which is exactly of the same form as the initial difference equation (\ref{diffeq1}), but using the dual variables. We call this the dual difference equation.

Let us comment on the role of the dual difference equation. If we consider only the difference equation given by $\mathsf O$, we remark that a formal solution can be multiplied by any $\ri \hbar$-periodic function in order to get another formal solution. By requiring that a solution is simultaneously a solution of the dual difference equation given by $\tilde {\mathsf O}$, we drastically reduce the set of eigenfunctions.
 Since the dual difference equation has the same form as the original difference equation, the small $\hbar$ expansion and the small $\hbar_D$ expansion of the eigenfunction should be closely related. We will use this argument when constructing the ``fully on-shell" eigenfunction from the  resummed WKB solutions.
 
 \subsection{WKB eigenfunction at small $\hbar$}
 
One possible approach to study the difference equation (\ref{diffeq1}) is to consider the small $\hbar$ regime, using the well known WKB ansatz:
\be
	\psi_{\rm WKB}(x)={\rm exp} \left (- \frac{1}{\ri \hbar} \sum_{n=0}^{\infty} S_n(x) (-\ri \hbar)^n  \right ).
\ee
Using this ansatz, for any $d \in \mathbb C$ we can write
\be
	\frac{\psi_{\rm WKB}(x + d \, \ri \hbar)}{ \psi_{\rm WKB}(x) } = {\rm exp} \left [- \frac{1}{\ri \hbar} \sum_{n=1}^{\infty}(-\ri \hbar)^n \left (\sum_{k=0}^{n-1} \frac{(-d)^{n-k}}{(n-k)!} S_k^{(n-k)}(x)  \right)   \right ].
\ee
Inserting this into the difference equation and expanding everything at small $\hbar$, we can recursively solve for $S_n'(x)$ order by order in $-\ri \hbar$. Then, we can integrate to obtain $S_n(x)$. Let us define $y(x)$ to be the solution of $W(x,y)=0$.
We find for the first orders
\be 
\ba
	S_0(x) = \int^x y(x)\rd x, \\
	S_1(x) =  \frac{1}{2}\log \frac{\partial y(x)}{\partial \kappa}.
\ea
\ee
The natural domain of $y(x)$ is not the $\mathbb C$-plane, but the mirror curve itself, which is a multi-sheeted cover of the plane. Since we consider the hyperelliptic case only, we have two sheets.\footnote{Strictly speaking, it is $\re^{y(x)}$ which is defined on a two-sheeted cover. } We need to take either branch of the function $y(x)$, which we also call $y(x)$ by abuse of notation. The two choices of the branch of $y(x)$ correspond to the WKB expansions of two independent solutions of the difference equation.
Let us define
\be
	X=\re^x.
\ee
At large $X$, we have
\be
\ba
	S_0(x) &= s_0(x)+S_0^{\rm inst}(X) = s_0(x)+ \int_\infty^X \tilde y(X') \frac{\rd X'}{X'}, \\
	S_1(x) &= s_1(x)+S_1^{\rm inst}(X),
\ea
\ee
where $s_0(x)$ is an order 2 polynomial, $s_1(x)$ is an order 1 polynomial, and $\tilde y(X)$ is equal to $y(x)$ minus the polynomial part in $x$ which appears in the large $X$ expansion. 
Both $s_0$ and $s_1$  are independent of $\kappa$.
It can be verified that the higher $S_n(x)$ are only functions of $X$: by this we mean that there is no polynomial in $x$ in the large $X$ expansion.
We use this remark and build the truncated WKB function
\be
	\label{truncWKB}
	\Psi_{\rm WKB}(X)=\psi_{\rm WKB}(x)\re^{-(-\ri \hbar)^{-1}s_0(x)-s_1(x)},
\ee
which only depends on $x$ through $X$. 

\subsection{Resummed WKB from recursion}

By adapting the manipulations done in \cite{acdkv}, we can resum the small $\hbar$ WKB expansion order by order in another expansion parameter, here $X^{-1}$. Let us define
\be
	q=\re^{\ri \hbar}.
\ee
 Shifting $x$ by $\ri \hbar$  in the eigenfunction $\psi_{\rm WKB}(x)$ correspond to multiplying $X$ by $q$ in the truncated function $\Psi_{\rm WKB}(X)$. 
The difference equation can be rewritten in terms of $\Psi_{\rm WKB}(X)$ only, by using the explicit forms of $s_0(x), s_1(x)$. In our hyperelliptic cases, it can be put in the form
\be
	\label{PsiEq1}
	\Psi_{\rm WKB}(q^{-1}X)-a(X)\Psi_{\rm WKB}(X)+b(X)\Psi_{\rm WKB}(q X)=0,
\ee
where $a(X)$ and $b(X)$ are rational functions of $X$, which also may depend on the moduli and parameters, as well as on $q^{1/2}$.
It is useful to perform a change of variables and use
\be
	\tilde X= \frac{X}{\kappa},
\ee
and
\be
	\tilde \Psi_{\rm WKB}(\tilde X)=\Psi_{\rm WKB}(X).
\ee
Using this, we find that (for the appropriate parametrization of the mirror curve) the difference equation takes the rather general form
\be
	(1+\tilde X^{-1}) \tilde \Psi_{\rm WKB}(\tilde X) -  \tilde \Psi_{\rm WKB}(q^{-1}\tilde X) +\frac{1}{\kappa^r} P[\tilde \Psi_{\rm WKB}( \tilde X) ]=0,
\ee
where $r$ is a strictly positive integer, and $P$ is the remaining part coming from the difference equation. This form suggests that we can solve this $q$-equation in a large $\kappa$ expansion. The ansatz we use is
\be
	\Psi_{\rm WKB}(\tilde X) = \Psi^{(0)}(\tilde X) \re^{\sum_{k=1}^{\infty} \phi_k(\tilde X)\kappa^{-k} }.
\ee
The leading part $\Psi^{(0)}(\tilde X)$ is universal, and is essentially a quantum dilogarithm:
\be
	 \Psi^{(0)}(\tilde X) = \prod_{N=0}^{\infty}(1+X^{-1}q^{-N-1})={\rm exp} \left ( \sum_{k=1}^{\infty}\frac{1}{k(1-q^k)}(-\tilde X)^k \right ).
\ee
To perform the recursion at large $\kappa$, we divide everything by $\Psi_{\rm WKB}(\tilde X)$ and use that
\be
	\label{gkfunctions}
\ba
	 \frac {\tilde \Psi_{\rm WKB}(q^{-1} \tilde X)}{ \tilde \Psi_{\rm WKB}( \tilde X)} &= (\tilde X^{-1}+1) \re^{\sum_{k=1}^{\infty} g_k(\tilde X)\kappa^{-k} }, \\
	  \frac {\tilde \Psi_{ \rm WKB}(q \tilde X)}{\tilde \Psi_{\rm WKB}( \tilde X)} &= \frac{1}{q^{-1}\tilde X^{-1}+1} \re^{-\sum_{k=1}^{\infty} g_k(q \tilde X)\kappa^{-k} }.
\ea
\ee
At each order in large $\kappa$, we get a linear equation determining $g_k( \tilde X)$ recursively. The recursion can be solved for $g_k(\tilde X)$, which are rational functions of $\tilde X$. The functions $\phi_k( \tilde X)$ are given by
\be
	g_k(\tilde X)=\phi_k(q^{-1} \tilde X)-\phi_k( \tilde X),
\ee
which can be formally solved in the following way:
\be
	\phi_k( \tilde X) = - \sum_{n=0}^{\infty}g_{k}(q^{-n}\tilde X).
\ee
This is especially useful if we work with an $\tilde X$-expanded form for $g_{k}(\tilde X)$, since
\be
	  \sum_{n=0}^{\infty}(q^{-n}\tilde X)^{-k}= \frac{\tilde X^{-k}}{1-q^k},
\ee
Also, we find that the larger $k$ is, the larger is the leading power of $\frac{1}{\tilde X}$ in the large $\tilde X$ expansion of $g_k(\tilde X)$.
In the end, after going back to the original variable $X$, we find the following structure:
\be
	\label{logVdef}
\ba
	\log V(X) &\equiv  \log  \frac {\Psi_{\rm WKB}(q^{-1} \tilde X)}{\Psi_{\rm WKB}( \tilde X)} \\
	&= \sum_{k=1}f_{k}(\kappa,{\bxi},q) X^{-k},
\ea
\ee
and so, formally,
\be
	\label{logWKB}
	\log \Psi_{\rm WKB}(X) = \sum_{k=1}\frac{f_{k}(\kappa,\bxi,q)}{q^k-1}X^{-k}.
\ee
In the above, we have collected the $\{ \xi_i \}_{i \geq 1}$ into the vector $\bxi$. 
The $f_{k}(\kappa,\bxi,q)$ are  polynomials in the variables $\kappa$ and $\xi_i$. If we expand this expression at small $\hbar$, we retrieve the large $X$ expansion of all the WKB corrections. So this expression is effectively a resummation of the small $\hbar$ WKB expansion.
To illustrate this procedure, we give here as examples the resummed WKB eigenfunctions for some cases which are associated to mirror curves of toric CY threefolds.

For the geometry called local $\mathbb P^2$, we have only one true modulus which is $\kappa$. The mirror curve and the WKB eigenfunction are
\be
	\label{P2mcwkb}
\ba
	W(x,y) &=\re^{x}+\re^{y}+\re^{-x-y}+\kappa, \\[0.2cm]
	%
	%
	\log \Psi_{\rm WKB}(X) &=\frac{\kappa }{(q-1) X}+\frac{\kappa ^2}{\left(2-2 q^2\right) X^2}+\frac{\frac{3}{q^{3/2}}-\kappa ^3}{\left(3-3 q^3\right) X^3}+\frac{\kappa  \left(\kappa ^3
   \left(-q^{5/2}\right)+4 q+4\right)}{4 q^{5/2} \left(q^4-1\right)
   X^4} \\
  & \quad +\frac{\kappa ^2
   \left(\kappa ^3-\frac{5 \left(q^2+q+1\right)}{q^{7/2}}\right)}{5 \left(q^5-1\right)
   X^5}-\frac{\kappa ^6 q^6+3 q^3-6 \kappa ^3 q^{3/2} (q+1) \left(q^2+1\right)+6}{6 q^6
   \left(q^6-1\right) X^6}+ \ldots
  \ea
\ee
For the geometry called local ${\mathbb P}^1 \times {\mathbb P}^1$, we have one true modulus which is $\kappa$, and one extra parameter $\xi_1$ which is a mass parameter. We rename it $m$. The mirror curve and the WKB eigenfunction are
\be
	\label{P1xP1mcwkb}
\ba
	W(x,y) &=\re^{x}+m\re^{-x}+\re^{y}+\re^{-y}+\kappa, \\[0.2cm]
	%
	%
	\log \Psi_{\rm WKB}(X) &=\frac{\kappa }{(q-1) X}+ \frac{-2 m q+\kappa ^2 q+2}{\left(2 q-2 q^3\right) X^2}+\frac{\kappa  \left(q \left(-3 m
   q+\kappa ^2 q+3\right)+3\right)}{3 q^2 \left(q^3-1\right) X^3} \\
  &\quad  -\frac{2 (m q-1) \left(q^2
   (m q-1)-2\right)+4 \kappa ^2 q \left(-m q^3+q^2+q+1\right)+\kappa ^4 q^4}{4 q^4
   \left(q^4-1\right) X^4}+ \ldots
\ea
\ee
For the geometry called the resolution of ${\mathbb C}^3 / {\mathbb Z}_5$, we have two  true moduli, $\kappa$ and an extra one  which is $\xi_1$. We rename this second one $\kappa_1$.  The mirror curve and the WKB eigenfunction are
\be
	\label{C3Z5mcwkb}
\ba
	W(x,y) &=\re^{x}+\re^{y}+\re^{-3x-y}+\kappa_1 \re^{-x}+\kappa, \\[0.2cm]
	%
	%
	\log \Psi_{\rm WKB}(X) &=\frac{\kappa }{(q-1) X}+\frac{\kappa
   ^2-2\kappa_1}{\left(2-2 q^2\right) X^2}+\frac{\kappa ^3-3 \kappa  \kappa_1}{3 \left(q^3-1\right) X^3} \\
   & \quad
   +\frac{\kappa ^4-4 \kappa ^2 \kappa_1+2 \kappa_1^2}{\left(4-4 q^4\right)
   X^4}+ \frac{-\kappa ^5+5 \kappa ^3\kappa_1-5 \kappa \kappa_1^2+\frac{5}{q^{5/2}}}{\left(5-5 q^5\right) X^5}\\
   & \quad 
   -\frac{-\left(\kappa ^2-2\kappa_1\right) \left(\kappa ^4-4 \kappa ^2\kappa_1+\text{$\kappa $1}^2\right)+\frac{6 \kappa }{q^{5/2}}+\frac{6 \kappa
   }{q^{7/2}}}{\left(6-6 q^6\right) X^6}
   +\ldots
  \ea
\ee
In each of these three cases, the polynomial part of the large $X$ WKB expansion is given by\footnote{This depends on the choice of branch for $y(x)$. Here we choose the one which reproduces (\ref{s0s1Ex}).}
\be
	\label{s0s1Ex}
	\frac{1}{-\ri \hbar }s_0(x)+s_1(x)  =  \frac{\ri}{2\hbar}x^2-\frac{1}{2}\left (\frac{2\pi}{\hbar} +1 \right ) x.
\ee
 
 
 \sectiono{The rational case}
 \label{sect3}
 
 In the following, we focus on the rational case, where $\hbar$ is given by $2\pi$ times a rational number. Using pole cancellation and modularity, we manage to write down an exact formula for a formal eigenfunction $\psi(x)$. 
 We will see that requiring modular invariance for this eigenfunction fixes all the true moduli, and $\psi(x)$ then becomes the ``fully on-shell" eigenfunction.
 The truncated WKB eigenfunction $\log \Psi_{\rm WKB}(X)$ is the only ingredient, but it comes with its modular dual which is invisible in the small $\hbar$ WKB expansion. As we will see, the resummation of the large $X$ expansion of $\log \Psi_{\rm WKB}$ can be down explicitly in the rational case.  
 
 \subsection{Pole cancellation and modular duality}
 \label{subsect31}
 When $\hbar$ is of the form
\be
	\hbar=2 \pi \frac{P}{Q},
\ee
for positive coprime integers $P$ and $Q$, the quantity $q=\re^{\frac{2 \pi \ri P}{Q}}$ is a root of unity:
\be
	q^Q=1.
\ee
 The formal solution (\ref{logWKB}) is ill-defined since it has poles when $k$ is a multiple of $Q$. We introduce a regulating parameter $\epsilon$ and consider the small $\epsilon$ expansion by setting
\be	
	\label{hbarepsilon}
	\hbar = 2 \pi \frac{P}{Q}+\epsilon.
\ee
We expand (\ref{logWKB}) in small $\epsilon$ by using
\be
	 \frac{1}{q^{\ell Q}-1} = -\frac{\ri}{\epsilon \ell Q}-\frac{1}{2}+{\mathcal O}(\epsilon),
\ee
and find
\be
	\label{pole1}
\ba
	\log \Psi_{\rm WKB}(X) &= -\frac{\ri}{\epsilon} \sum_{\ell=1}^{\infty} \frac{f_{\ell Q}(\kappa,\bxi,q)}{\ell Q}X^{-\ell Q}+ {\mathcal O}(1).
\ea 
\ee
As it is, the naive resummation of the WKB expansion given by $\log \Psi_{\rm WKB}$ is singular at rational $\hbar/2\pi$. We conclude that it has to be corrected by something which
\vspace{0.2cm}

	1) is non-pertubative at small $\hbar$, so that it is invisible in the small $\hbar$ WKB expansion,
\vspace{0.2cm}	

	2) cancels the poles in the rational case.
\vspace{0.2cm}

\noindent
Also, we have not taken into account the modular duality structure outlined in the previous section. Indeed, our point of view was to start with the small $\hbar$ WKB resummation of the eigenfunction. However, we could have equally well started from the dual equation (\ref{dualDiffEq1}) also satisfied by the ``fully on-shell" eigenfunction, and consider its small $\hbar_D$ WKB expansion. 
By doing the same recursive procedure, we would end up with a very similar expression for the truncated dual WKB eigenfunction
\be
	 \log \Psi_{\rm WKB}^D (X) = \sum_{k=1}\frac{f_{k}(\tilde \kappa, \tilde \bxi,q_D)}{{q_D}^k-1}X_D^{-k},
 \ee
 where  $X_D=\re^{x_D}=\re^{\frac{2\pi}{\hbar}x}$, $\hbar_D=4\pi^2/\hbar$ and $q_D=\re^{\ri \hbar_D}=\re^{ \frac{2\pi Q}{P} }$. The $f_k$ are precisely the same polynomials as in (\ref{logWKB}), since the dual equation is of the same form as the initial equation.
So we would expect an eigenfunction which is invariant under the exchanges $\hbar \leftrightarrow \frac{4\pi^2}{\hbar}$ and $(\kappa,\bxi) \leftrightarrow (\tilde \kappa, \tilde \bxi)$. 
Following what is suggested in \cite{KP}, let us add its dual to the resummed WKB, which is a non-perturbative contribution at small $\hbar$:
\be
\ba
	\label{logWKB2}
	\log \Psi(X) &= \log \Psi_{\rm WKB}(X) +  \log \Psi_{\rm WKB}^D (X) \\
	& = \sum_{k=1}\frac{f_{k}(\kappa,\bxi,q)}{q^k-1}X^{-k}+\sum_{k=1}\frac{f_{k}(\tilde \kappa, \tilde \bxi,q_D)}{{q_D}^k-1}X_D^{-k}.
\ea
\ee
 The dual part also has poles when $k=\ell P$ for integer $\ell$. Using (\ref{hbarepsilon}) and expanding at small $\epsilon$, we obtain
\be
\ba
	\log \Psi^D_{\rm WKB}(X) &= \frac{\ri}{\epsilon} \frac{P}{Q} \sum_{\ell = 1}^{\infty} \frac{f_{\ell P}(\tilde \kappa, \tilde \bxi,q_D)}{\ell Q}X^{-\ell Q} +O(1).
\ea
\ee
In the full expression (\ref{logWKB2}), this pole cancels with the corresponding pole in (\ref{pole1}) \emph{if} the following condition is fulfilled:
\be
	\label{cancelCond1}
	P \, f_{\ell P}(\tilde \kappa, \tilde \bxi,q_D)=Q \, f_{\ell Q}( \kappa,  \bxi,q)
\ee
for all positive integers $\ell$. This defines relations
\be
	\label{kappaRel1}
	\tilde \kappa(\kappa, \bxi;\hbar), \qquad {\rm and} \qquad  \tilde \bxi(\kappa,\bxi;\hbar).
\ee
Since the $f_k$ are polynomials in $\kappa$ and $\xi_k$, these relations are algebraic \emph{at fixed rationsl $\hbar$}. The system of equations (\ref{cancelCond1}) seems strongly overdetermined, but nevertheless we find that there actually are solutions as a consequence of the form of the $f_k$. Some examples can be found below.
The ``fully on-shell" values of $\kappa$ and $\boldsymbol \xi$ depend on $\hbar$, so we should write $\kappa(\hbar), \boldsymbol \xi(\hbar)$ and $\tilde \kappa(\kappa(\hbar),  \bxi(\hbar) ;\hbar), \tilde \bxi(\kappa(\hbar),\bxi(\hbar) ;\hbar)$. Once these relations are fixed, our claim is that (\ref{logWKB2}) is the full non-pertrubatively completed WKB eigenfunction in the large $X$ expansion, and can be used to build the ``fully on-shell" eigenfunctions. 
 
 Let us remark quickly that, in contrast with \cite{KP}, we did not make use of the so called quantum mirror map, or quantum A periods. In \cite{KP} (and also in the setup of the TS/ST correspondence of \cite{ghm, cgm, mz1}), the quantum mirror map fixes the relations between the moduli/parameters $\kappa$ and $\xi_i$ and their duals $\tilde \kappa$ and $\tilde \xi_i$. Here, we impose these relations in the rational case using pole cancellation. This is less general but in some sense more natural and straightforward from the point of view of the difference equation.
 
 \subsection{Finite contribution}
 
Requiring cancellation of poles and modular duality is what motivated us to write expression (\ref{logWKB2}). Let us now work out the finite terms at $\hbar= 2\pi P/Q$. We insist that when varying $\hbar$, we should also vary the modulus $\kappa$ (corresponding to the the eigenvalue) as well as all the other true moduli among the $\bxi$.
By using that $q^{k+\ell Q}=q^k$, the first part of (\ref{logWKB2}) gives
\be
	\label{finite1}
\ba
	 \sum_{k=1}\frac{f_{k}(\kappa,\bxi,q)}{q^k-1}X^{-k} &= -\frac{\ri}{\epsilon} \sum_{\ell=1}^{\infty} \frac{f_{\ell Q}(\kappa,\bxi,q)}{\ell Q}X^{-\ell Q} \\
		& \qquad + \sum_{\ell=0}^{\infty}  \frac{-\ri \partial_\hbar f_{\ell Q}(\kappa,\bxi,q)}{\ell Q}X^{-\ell Q}-\frac{1}{2} \sum_{\ell=0}^{\infty}   f_{\ell Q}(\kappa,\bxi,q) X^{-\ell Q}\\	
		& \qquad - \ri \kappa' \frac{\partial}{\partial \kappa} \sum_{\ell=0}^{\infty}  \frac{f_{\ell Q}(\kappa,\bxi,q)}{\ell Q}X^{-\ell Q}
		- \ri \left ( \bxi'  \cdot\frac{\partial}{\partial \bxi} \right)\sum_{\ell=0}^{\infty}  \frac{f_{\ell Q}(\kappa,\bxi,q)}{\ell Q}X^{-\ell Q}
		\\
		& \qquad + \sum_{k=1}^{Q-1} \frac{1}{q^k-1} \sum_{\ell=0}^{\infty} f_{\ell Q+k}(\kappa,\bxi,q)X^{-\ell Q-k} \\
		& \qquad+ {\mathcal O}(\epsilon),
\ea 
\ee
where $\partial_\hbar f_k(\kappa,\bxi,q)=\ri q \partial_q f_k(\kappa,\bxi,q)$. We have denoted $\kappa'$ and $\bxi'$ the $\hbar$ derivatives of $\kappa$ and $\bxi$. Every term in the sum above can be written using the function $\log V(X)$ defined in (\ref{logVdef}), which is finite for any value of $\hbar$. The function $\log V(X)$ is for the moment only known as a large $X$ expansion. However, as we will see below, it can be obtained exactly for any integers $Q,P$. For $a,k$ integers we have that $\frac{1}{Q}\sum_{m=0}^{Q-1}q^{m(a-k)}=\delta_{0,(a-k) \,\, {\rm mod} \,\, Q}$. We define for $k=1,2,\ldots,Q$,
\be
	\label{defphik}
	\varphi_k(X) \equiv \sum_{\ell=0}^{\infty} f_{\ell Q+k}(\kappa,\bxi,q) X^{-\ell Q-k}=\frac{1}{Q} \sum_{m=0}^{Q-1} \log V(q^m X)q^{m k}.
\ee
The instance with $k=Q$ can also be written as
\be
	\varphi_Q(X) \equiv \sum_{\ell=1}^{\infty} f_{\ell Q}(\kappa,\bxi,q) X^{-\ell Q}=\frac{1}{Q} \sum_{m=0}^{Q-1} \log V(q^m X).
\ee
The function $\varphi_Q$ also appears in the following combination:
\be
	 \sum_{\ell=1}^{\infty} \frac{f_{\ell Q}(\kappa,\bxi,q)}{\ell Q}X^{-\ell Q} = -\int_{\infty}^X \varphi_Q(X')\frac{\rd X'}{X'}.
\ee
So (\ref{finite1}) becomes
\be
\ba
	 \sum_{k=1}\frac{f_{k}(\kappa,\bxi,q)}{q^k-1}X^{-k} &= \frac{\ri}{ \epsilon } \int_{\infty}^X \varphi_Q(X')\frac{\rd X'}{X'} \\
	& \qquad + \int_{\infty}^X \ri \partial_{\hbar} \varphi_Q(X')\frac{\rd X'}{X'} 
	+ \ri \kappa' \int_{\infty}^X  \partial_{\kappa} \varphi_Q(X')\frac{\rd X'}{X'} \\
	&+  \ri \bxi'  \cdot \int_{\infty}^X  \partial_{\bxi} \varphi_Q(X')\frac{\rd X'}{X'} 
	-\frac{\varphi_Q(X)}{2}   +\sum_{k=1}^{Q-1} \frac{\varphi_k(X)}{q^k-1} 
	  + {\mathcal O}(\epsilon).
\ea
\ee
The function $\partial_{\hbar} \varphi_Q(X')$ can be obtained through
\be
	\label{dhphiQ}
	\partial_{\hbar} \varphi_Q(X) = \frac{1}{Q} \sum_{m=0}^{Q-1} \left. \frac{\partial_\hbar V}{V} \right |_{X \rightarrow q^m X},
\ee
whereas the functions $\partial_\kappa \varphi_Q$ and $\partial_{\bxi} \varphi_Q$ can be obtained by direct differentiation of $\varphi_Q$.

Let us now look at the expansion of the dual part. We find
\be
	\label{finite2}
\ba
	\sum_{k=1}\frac{f_{k}(\tilde \kappa, \tilde \bxi,q_D)}{{q_D}^k-1}X_D^{-k} &= \frac{\ri}{\epsilon} \frac{P}{Q} \sum_{\ell = 1}^{\infty} \frac{f_{\ell P}(\tilde \kappa, \tilde \bxi,q_D)}{\ell Q}X^{-\ell Q} \\
		& \qquad  +  \frac{\ri}{2\pi} \left ( \sum_{\ell=1}^{\infty}\frac{f_{\ell P}(\tilde \kappa, \tilde \bxi,q_D)}{\ell Q}X^{-\ell Q} + x   \sum_{\ell=1}^{\infty}f_{\ell P}(\tilde \kappa, \tilde \bxi,q_D) X^{-\ell Q}  \right )  \\
		& \qquad + \sum_{\ell=1}^{\infty} \frac{-\ri \partial_{\hbar_D} f_{\ell P}(\tilde \kappa, \tilde \bxi,q_D) }{\ell P}  X_D^{-\ell P} -\frac{1}{2} \sum_{\ell=1}^{\infty}   f_{\ell P}(\tilde \kappa, \tilde \bxi,q_D) X^{-\ell Q} \\
		& \qquad +\ri \frac{P}{Q} \left (  {\tilde \kappa}'  \sum_{\ell=1}^{\infty} \frac{\partial_{\tilde \kappa}f_{\ell P}(\tilde \kappa, \tilde \bxi,q_D) }{\ell Q}  X^{-\ell Q}
		+  \left ( {\tilde \bxi}' \cdot \frac{\partial}{\partial \tilde \bxi} \right ) \sum_{\ell=1}^{\infty} \frac{f_{\ell P}(\tilde \kappa, \tilde \bxi,q_D) }{\ell Q} X^{-\ell Q} \right ) \\
		& \qquad +\sum_{k=1}^{P-1}\frac{1}{q_D^k-1} \sum_{\ell=0}^{\infty} f_{\ell P+ k}(\tilde \kappa,\tilde \bxi,q_D){X_D}^{-\ell P-k}+ {\mathcal O}(\epsilon).
\ea
\ee
Explicitly, $X_D=X^{Q/P}$. Also, $\tilde \kappa$ and $\tilde \bxi$ have several sources of $\hbar$ dependance, and we have denoted their total derivative w.r.t. $\hbar$ by $\tilde \kappa'$ and $\tilde \bxi'$.
As before, we can write all the contributions in terms of a unique function $\log V_D(X_D)$, which is defined as
\be
	\log V_D(X_D) \equiv  \sum_{k=1}f_{k}(\tilde \kappa, \tilde \bxi,q_D) X_D^{-k}.
\ee
This is basically $\log V(X)$ where we replaced all the variables by their duals. 
As before, we define for $k=1,2,\ldots,P$,
\be
	\tilde \varphi_k(X_D) \equiv \sum_{\ell=0}^{\infty} f_{\ell P+k}(\tilde \kappa,\tilde \bxi,q_D) X_D^{-\ell P-k}=\frac{1}{P} \sum_{m=0}^{P-1} \log V_D(q_D^m X_D)q_D^{m k}.
\ee
Using (\ref{cancelCond1}), we see that the special case $k=P$ is related to $\varphi_Q$:
\be
\ba
	\tilde \varphi_P(X_D) =\frac{Q}{P} \varphi_{Q}(X).
\ea
\ee
Now that we have performed the variation with respect to $\hbar$, it is considered to be fixed in what follows. The relation between parameters and their duals are the algebraic ones (\ref{kappaRel1}) at fixed $\hbar$.  So we can write
\be
\ba
	 \sum_{\ell=1}^{\infty} \frac{\partial_{\tilde \kappa}f_{\ell P}(\tilde \kappa, \tilde \bxi,q_D) }{\ell Q}  X^{-\ell Q}
	 &= \frac{Q}{P} (  \partial_{\tilde \kappa}  \kappa \,\, \partial_{\kappa}+ \partial_{\tilde \kappa}  \bxi \cdot  \partial_{\bxi}) \sum_{\ell=1}^{\infty} \frac{f_{\ell Q}( \kappa,  \bxi,q)}{Q \ell}X^{-\ell Q} \\
	  &=  -\frac{Q}{P} \partial_{\tilde \kappa} \kappa \int_{\infty}^{X} \partial_{\kappa} \varphi_Q(X') \frac{\rd X'}{X'}-\frac{Q}{P} \partial_{\tilde \kappa} \bxi \cdot \int_{\infty}^{X} \partial_{\bxi} \varphi_Q(X') \frac{\rd X'}{X'},
\ea
\ee
and similarly for the term with the derivative w.r.t. $\bxi$. These are exactly the kind of terms appearing in the expansion of the first part.
 So (\ref{finite2}) becomes
\be
\ba
	\sum_{k=1}\frac{f_{k}(\tilde \kappa, \tilde \bxi,q_D)}{{q_D}^k-1}X_D^{-k} & = 
	-\frac{\ri}{ \epsilon } \int_{\infty}^X \varphi_Q(X')\frac{\rd X'}{X'} \\
	&\qquad + \frac{\ri Q}{2\pi P} \left (-\int_{\infty}^X \varphi_Q(X')\frac{\rd X'}{X'}+x \varphi_{Q}(X) \right ) \\
	& \qquad + \int_{\infty}^{X_D} \ri \partial_{\hbar_D} \tilde \varphi_P(X')\frac{\rd X'}{X'} -\frac{1}{2} \frac{Q}{P} \varphi_Q(X) \\
	& \qquad - \ri (\tilde \kappa' \partial_{\tilde \kappa} \kappa+\tilde \xi'_i \partial_{\tilde{\xi_i}}\kappa ) \int_{\infty}^{X}  \partial_{\kappa}   \varphi_Q(X')\frac{\rd X'}{X'}  \\
	& \qquad -  \ri (\tilde \kappa' \partial_{\tilde \kappa} \bxi+\tilde \xi'_i \partial_{\tilde{\xi_i}}\bxi ) \cdot \int_{\infty}^{X}  \partial_{\bxi}   \varphi_Q(X')\frac{\rd X'}{X'}  
	 \\
	& \qquad    +\sum_{k=1}^{Q-1} \frac{\varphi_k(X)}{q^k-1} 
	  + {\mathcal O}(\epsilon)
\ea
\ee
Finally, by adding this to what we obtained previously, we can write down the finite part of the full non-perturbative WKB eigenfunction (\ref{logWKB2}) in the rational case:
\be
	\label{logPsiRational}
\ba
	\log \Psi(X) &= \frac{\ri Q}{2\pi P} \left (-\int_{\infty}^X \varphi_Q(X')\frac{\rd X'}{X'}
	+\lambda \int_{\infty}^X \partial_{\kappa}\varphi_Q(X')\frac{\rd X'}{X'}
	+{\boldsymbol \lambda}_\xi \cdot \int_{\infty}^X \partial_{\bxi}\varphi_Q(X')\frac{\rd X'}{X'}+x \varphi_{Q}(X) \right )  \\
	& \qquad -\frac{1}{2}\left ( 1+ \frac{Q}{P} \right ) \varphi_Q(X)
		 + \int_{\infty}^{X} \ri \partial_\hbar \varphi_Q(X') \frac{\rd X'}{X'}+\int_{\infty}^{X_D} \ri \partial_{\hbar_D} \tilde \varphi_P(X_D') \frac{\rd X_D'}{X_D'} \\
		& \qquad + \sum_{k=1}^{Q-1} \frac{\varphi_k(X)}{q^k-1}+ \sum_{k=1}^{P-1} \frac{\tilde \varphi_k(X_D)}{{q_D}^k-1}.
\ea
\ee
Here we have collected in $\lambda$ and ${\boldsymbol \lambda}_\xi$ all the terms in front of the corresponding integrals. These terms cannot be determined directly by our method, and we will use monodromy invariance of the eigenfunction to fix them.
Let us also notice that in the first line in the parenthesis, the first and last term can be put together using integration by part, to give
$\int x  \rd \varphi_Q(x)$. So this first line corresponds to the integral of what is called the ``deformed symplectic potential" for the special case studied in \cite{ks}, with deformation parameters $\lambda$ and ${\boldsymbol \lambda}_\xi$.

Once we know how to build $\log V$ and $\log V_D$ and their $\hbar$ derivative, everything is exactly determined. We will construct them in the following subsection.
 
 \subsection{Exact expressions for the building blocks}

In this section, we present the method to compute exactly the various functions appearing in (\ref{logPsiRational}) in the rational case. As we will see, since we are in the hyperelliptic case, a certain product of $2 \times 2$ matrices will be crucial. We expect something similar for more general cases, with matrices of larger size. The method presented here can be seen as a kind of generalization of the manipulations done in \cite{hkt,hsx}.

We recall the definition of $V(X)$:
\be
	V(X)=\frac{ \Psi_{\rm WKB}(q^{-1}X) }{ \Psi_{\rm WKB}(X) }.
\ee
The difference equation (\ref{PsiEq1}), can be rewritten for $V(X)$ as
\be
	\label{VVab}
	V(X)V(q X)-a(X) V(q X)+b(X)=0,
\ee
where we remind that $a(X)$ and $b(X)$ are rational functions of $X$.
The key feature of the rational case is that this equation can be solved algebraically (using $q^Q=1$). To proceed, we shorten the notation by using $v_k = V(q^k X)$. The label $k$ of $v_k$ is thus defined modulo $Q$. The previous equation can be shifted, which gives the closed system of $Q$ quadratic equations for the $Q$ variables $v_k$, where $k=0,1,\ldots,Q-1$:
\be
	\label{vkvk1}
	v_k v_{k+1}-a(q^k X) v_{k+1}+b(q^k X)=0, \qquad \qquad k=0,1,\ldots,Q-1.
\ee
To efficiently solve this system, we proceed by recursion. We define $a^{(k)}(X)$ and $b^{(k)}(X)$ through the following relations
\be
	\label{ABvprods}
\ba
	 \, & a^{(1)}(X) =a(X),\\
	 & b^{(1)}(X) =b(X), \\
	 &v_0 v_1 \cdots v_{k}-a^{(k)}(X) v_{k}+b^{(k)}(X) =0.
\ea
\ee
The next term is obtained by multiplying the last line by $v_{k+1}$, and using (\ref{vkvk1}):
\be
\ba
	0 &= v_0 v_1 \cdots v_{k} v_{k+1} -a^{(k)}(X) v_{k}v_{k+1}+b^{(k)}(X)v_{k+1} \\
		&= v_0 v_1 \cdots v_{k+1} -[a(q^k X) a^{(k)}-b^{(k)}(X)]v_{k+1}+b(q^{k}X)a^{(k)}(X).
\ea
\ee
From this we read out the relation
\be
	\begin{pmatrix}
		a^{(k+1)}(X) \\
		b^{(k+1)}(X)
	\end{pmatrix}
	=
	\begin{pmatrix*}[r]
		a(q^k X) & \,\,  -1 \\
		b(q^k X)  & 0 
	\end{pmatrix*}
	\begin{pmatrix}
		a^{(k)}(X) \\
		b^{(k)}(X)
	\end{pmatrix}.
\ee
This recursion can be easily solved, and we find, for example for $k=Q$,
\be
	\label{relQm1}
	\begin{pmatrix}
		a^{(Q)}(X) \\
		b^{(Q)}(X)
	\end{pmatrix}
	=
	{\mathcal M}(X)
	\begin{pmatrix}
		1 \\
		0
	\end{pmatrix}
	=\begin{pmatrix}
		{\mathcal M}_{11}(X) \\
		{\mathcal M}_{21}(X)
	\end{pmatrix},
\ee
where the matrix
\be
	\label{Mdef}
{\mathcal M}(X)=
	\prod_{k=1}^{Q}
	\begin{pmatrix*}[r]
		a(q^{-k} X) & \,\,  -1 \\
		b(q^{-k} X)  & 0 
	\end{pmatrix*}
\ee
is defined such that the product is ordered from left to right as $k$ increases.
Using that 
\be
	\label{MqM}
	{\mathcal M}(q X) = 
	\begin{pmatrix*}[r]
		a(X) & \,\,  -1 \\
		b(X)  & 0 
	\end{pmatrix*}
	{\mathcal M}(X)
	\begin{pmatrix*}[r]
		a( X) & \,\,  -1 \\
		b(X)  & 0 
	\end{pmatrix*}^{-1},
\ee
we get
\be
	\begin{pmatrix}
		a^{(Q)}(qX) \\
		b^{(Q)}(qX)
	\end{pmatrix}
	=\begin{pmatrix}
		{\mathcal M}_{22}(X)-a(X){\mathcal M}_{12}(X) \\
		b(X){\mathcal M}_{12}(X)
	\end{pmatrix}.
\ee
Let us define
\be
	\Pi v = v_0 v_1 \cdots v_{Q-1},
\ee
which is invariant under $q$-shifts. We obtain from (\ref{ABvprods}) (for $k=Q$):
\be
	\begin{cases}
	v_0(\Pi v- a^{(Q)}(X))+b^{(Q)}(X)=0, \\
	v_1(\Pi v- a^{(Q)}(qX))+b^{(Q)}(qX)=0.
	\end{cases}
\ee
The second line is just the $q$-shift of the first. Using the expressions of $a^{(Q)}$ and $b^{(Q)}$ in terms of the entries of $\mathcal M$ and then (\ref{vkvk1}), we can rewrite this system as
\be
	\label{v0system}
	\begin{cases}
	v_0(\Pi v - {\mathcal M}_{11}(X))+ {\mathcal M}_{21}(X)=0, \\
	-(\Pi v - {\mathcal M}_{22}(X))-v_0  {\mathcal M}_{12}(X)=0.
	\end{cases}
\ee
This system can also be rewritten in matrix form:
\be
	\begin{pmatrix}
		{\mathcal M}_{11}(X)-\Pi v & {\mathcal M}_{21}(X) \\
		{\mathcal M}_{12}(X) & {\mathcal M}_{22}(X)-\Pi v
	\end{pmatrix}
	\begin{pmatrix}
		-v_0 \\
		1
	\end{pmatrix}
	=
	\begin{pmatrix}
		0 \\
		0
	\end{pmatrix},
\ee
which has solutions only if
\be
	0 = {\rm det}({\mathcal M}^{\rm T}- \Pi v \, {\bf 1})= {\rm det}({\mathcal M}- \Pi v \, {\bf 1}).
\ee
We conclude that $\Pi v$ is an eigenvalue of the matrix $\mathcal M$(X):
\be
	\Pi v = v_0 v_1 \cdots v_{Q-1} =  \frac{ {\rm tr}{\mathcal M}(X) \pm \sqrt{\Delta(X)} }{2},
\ee
where
\be
	\Delta(X) =  ({\rm tr}{\mathcal M}(X))^2 - 4 \,\, {\rm det}{\mathcal M}(X).
\ee
Both ${\rm det}{\mathcal M}(X)$ and ${\rm tr}{\mathcal M}(X) $ are invariant under $q$-shifts (see (\ref{MqM})) and so depend on $X$ through $X^Q$. The function $v_0$ can be found for example using the second line of (\ref{v0system}):
\be
	v_0 = \frac{ \Pi v - {\mathcal M}_{22}(X) }{- {\mathcal M}_{12}(X)} =  \frac{ {\mathcal M}_{11}(X)- {\mathcal M}_{22}(X) \pm \sqrt{\Delta(X)} }{-2 {\mathcal M}_{12}(X)}.
\ee
The other $v_k$ can be obtained by $q$-shifting $v_0$.
We thus provided the solution of the $q$-equation (\ref{VVab}) for all the rational cases: the solution is encoded in the matrix $\mathcal M(X)$ which can be obtained by  (\ref{Mdef}), i.e. a product of $Q$ matrices which are $q$-shifted. Finally, we find the following results for $\log V(X)$ and $\varphi_Q(X)$:
\be
\ba
	\log V(X) &= \log \left ( \frac{ {\mathcal M}_{11}(X)- {\mathcal M}_{22}(X) \pm \sqrt{\Delta(X)} }{-2 {\mathcal M}_{12}(X)} \right ), \\
	\varphi_Q(X) &= \frac{1}{Q} \log \left ( \frac{ {\rm tr}{\mathcal M}(X) \pm \sqrt{\Delta(X)} }{2} \right ). \\
\ea
\ee

The dual quantities $\log V_D(X_D)$ and $\tilde \varphi_{P}(X_D) $ can be obtained by exchanging $Q$ and $P$ and replacing all the variables by their duals $X_D$, $\tilde \kappa$ and $\tilde {\bxi}$. This means redefining $a(X)$ and $b(X)$ since they have implicit dependance on $\kappa$, $\bxi$ and perhaps $q$. For convenience, we write the results here:
\be
\ba
	\tilde {\mathcal M}(X_D) &=\prod_{k=1}^{P}
	\begin{pmatrix*}[r]
		a_D(q^{-k} X_D) & \,\,  -1 \\
		b_D(q^{-k} X_D)  & 0 
	\end{pmatrix*}, \\[0.2cm]
	\Delta_D(X_D) &= ({\rm tr} \tilde {\mathcal M}(X_D))^2-4 \, {\rm det} \tilde {\mathcal M}(X_D), \\[0.2cm]
	\log V_{D}(X_D) &= \log \left ( \frac{ \tilde {\mathcal M}_{11}(X_D)- \tilde {\mathcal M}_{22}(X_D) \pm \sqrt{\Delta_D(X_D)} }{-2 \tilde {\mathcal M}_{12}(X_D)} \right ), \\
	\tilde \varphi_{P}(X_D) &= \frac{1}{P} \log \left (  \frac{ {\rm tr} \tilde {\mathcal M}(X_D) \pm \sqrt{\Delta_D(X_D)} }{2} \right ).
\ea
\ee

All these functions are determined up to the sign in front of the square root. This freedom of choice corresponds to the branch choice of $y(x)$ in the WKB method of the previous section. As we will see, in the final eigenfunctions both choices appear in a symmetric way.

By now, the only ingredients appearing in (\ref{logPsiRational}) which have not been explicitly constructed are $\partial_\hbar \varphi_Q(X)$ and its dual. They cannot be obtained by simply taking $\hbar$ derivatives of $\varphi_Q$ and $\tilde \varphi_P$ because we do not know their explicit $\hbar$ dependance as exact functions. We only know their $\hbar$ dependance as a large $X$ expansion, or an $\hbar$ dependant algorithm to build them in the rational case. To find an expression for $\partial_\hbar \varphi_Q(X)$ in the rational case, we basically perform a first order WKB expansion but around $\hbar=2\pi P/Q$ instead of $\hbar=0$. 
In order to do this, let us take a total $\hbar$-derivative of equation (\ref{VVab}), which is valid for any $\hbar$:
\be
\ba
	0 &= \partial_\hbar V(X)V(qX)+V(X)\partial_\hbar V(q X)+\ri V(X) \partial_x V(q X) \\
	& \qquad -\partial_\hbar a(X)V(q X)-a(X)\partial_\hbar V(q X)-\ri a(X)\partial_x V(q X)+\partial_\hbar b(X).
\ea
\ee
This is the  $q$-equation obeyed by the first derivative of $V$. It can be solved in the rational case.
Using the notation
\be
	\delta(X) = \frac{ \partial_\hbar V}{V}(X),
\ee
this can be rewritten as
\be
	\label{hbarpert1}
	0=\delta(q X)+\delta(X) \left ( 1-\frac{a(X)}{V(X)} \right )^{-1}+\left (\ri \partial_x \log V(q X)- \left ( 1-\frac{a(X)}{V(X)} \right )^{-1} \frac{\partial_\hbar a(X)}{V(X)}+\frac{\partial_\hbar b(X)}{b(X)}  \right )
\ee
Similarly as before, let us introduce the notations
\be
\ba
	\delta_k & = \delta(q^k X), \\
	v_k & = V(q^k X),
\ea
\ee
and
\be
\ba
	\alpha(X) & =  -\left ( 1-\frac{a(X)}{v_0} \right )^{-1} \\
	\beta(X) & = -\left (\ri \partial_x \log v_1- \left ( 1-\frac{a(X)}{v_0} \right )^{-1} \frac{\partial_\hbar a(X)}{v_0}+\frac{\partial_\hbar b(X)}{b(X)}  \right ).
\ea
\ee
Both $\alpha(X)$ and $\beta(X)$ are of the form
\be
	({\text{rational of }}X) \pm ({\text{rational of }}X) \times \sqrt{\Delta(X)},
\ee
 Every further manipulations will leave invariant this structure, so the final result will also be of this form.
Equation (\ref{hbarpert1}) can be written as
\be
	0=\delta_1 - \alpha(X) \delta_0-\beta(X),
\ee
which can be treated similarly as in the previous section. It is in principle even simpler since it is a polynomial of order 1 in the $\delta_k$, instead of order 2. By recursion,
\be
\ba
	\, & \alpha^{(1)}(X) =\alpha(X),\\
	& \beta^{(1)}(X) =\beta(X), \\
	&  \delta_k  - \alpha^{(k)}(X)\delta_0-\beta^{(k)}(X) =0.
\ea
\ee
After shift,
\be
\ba
	0 &= \delta_{k+1}  - \alpha^{(k)}(qX)\delta_1-\beta^{(qk)}(X),  \\
	  &= \delta_{k+1}  - \alpha^{(k)}(qX) \alpha(X) \delta_0-(\beta^{(k)}(qX)+\alpha^{(k)} \beta(X)), 
\ea
\ee
from which we read out
\be
	\begin{pmatrix}
		\alpha^{(k+1)}(X) \\
		\beta^{(k+1)}(X)
	\end{pmatrix}
	=
	\begin{pmatrix*}[r]
		\alpha(X) & \,\,  0 \\
		\beta(X)  & 1
	\end{pmatrix*}
	\begin{pmatrix}
		\alpha^{(k)}(qX) \\
		\beta^{(k)}(qX)
	\end{pmatrix}.
\ee
From this, we get
\be
	\begin{pmatrix}
		\alpha^{(Q)}(X) \\
		\beta^{(Q)}(X)
	\end{pmatrix}
	=
	{\mathcal A}(X)
	\begin{pmatrix}
		1 \\
		0
	\end{pmatrix},
\ee
where
\be
	{\mathcal A}(X)=
	\prod_{k=0}^{Q-1}
	\begin{pmatrix*}[r]
		\alpha(q^kX) & \,\,  0 \\
		\beta(q^kX)  & 1
	\end{pmatrix*}.
\ee
Again, the product is ordered from left to right as $k$ increases
Since we have $\delta_Q=\delta_0$, we end up with
\be
\ba
	0=\delta_0-\alpha^{(Q)}(X) \delta_0-\beta^{(Q)}(X),
\ea
\ee
which is solved by
\be
	 \delta_0  \equiv \frac{\partial_\hbar V}{V}(X) = \frac{\beta^{(Q)}(X)}{1-\alpha^{(Q)}(X)}.
\ee
From the recursion (or  its solution given by the matrix $\mathcal A$), it is easily seen that
\be
	\alpha^{(k)}(X) = \prod_{\ell=0}^{k-1}\alpha(q^\ell X),
\ee
which means that for $k=Q$, it is invariant under $q$-shifts. Also, we have
\be
	\beta^{(Q)}(X) = \sum_{k=0}^{Q-1}\beta(q^k X) \alpha^{(Q-1-k)}(q^{k+1}X),
\ee 
where we used the convention $\alpha^{(0)}(X)=1$.
Finally, according to (\ref{dhphiQ}),
\be
	\partial_{\hbar} \varphi_Q(X) = \frac{1}{Q}\sum_{k=0}^{Q-1}\frac{\beta^{(Q)}(q^kX)}{1-\alpha^{(Q)}(q^k X)}= \frac{1}{Q}\frac{\sum_{k=0}^{Q-1} \beta^{(Q)}(q^kX)}{1-\alpha^{(Q)}(X)} ,
\ee
where we used invariance of $\alpha^{(Q)}(qX)$ under $q$-shifts.
We can change the order of summation, to obtain the following form, which is more useful in actual computations:
\be
	\partial_{\hbar} \varphi_Q(X) =\frac{1}{Q(1-\alpha^{(Q)}(X))} \sum_{N=0}^{Q-1} \beta(q^N X) \left ( \sum_{k=0}^{Q-1} \alpha^{(k)}(q^{N+1}X) \right ).
\ee
The dual quantity $\partial_{\hbar_D} \tilde \varphi_P(X_D)$ is of course built in the same way, where we exchange $Q$ and $P$ and use the dual quantities everywhere.
In principle, we now have all the ingredients to write down (\ref{logPsiRational}) exactly. 

Let us remark that all these ingredients are functions which are multivalued (the sign ambiguity in front of the square-root). So we must consistently choose a branch of these functions. As we will see in the final result, both choices will contribute.

\subsection{Relations between the parameters}
\label{sectRels}

We saw in section \ref{subsect31} that we need some conditions on the functions $f_k$ for pole cancellation, which translate into relations between $\kappa, \bxi$ and $\tilde \kappa,\tilde \bxi$. Here we make this relation more explicit for the rational case, and give some examples. As we have already seen, condition (\ref{cancelCond1}) can be rewritten as the functional relation 
\be
	Q \varphi_{Q}(X,\kappa,\bxi)= P \tilde \varphi_P(X_D, \tilde \kappa,\tilde \bxi),
\ee
valid for all $X$.
Using results from the previous subsection, this is equivalent to
\be
	\label{trpmDelta}
	  {\rm tr}{\mathcal M}(X) \pm \sqrt{\Delta(X)} =  {\rm tr}{\mathcal M_D}(X_D) \pm \sqrt{\Delta_D(X_D)}.
\ee
A necessary condition for this to hold is the equality of the traces for all $X$
\be
	{\rm tr}{\mathcal M}(X) =  {\rm tr}{\mathcal M_D}(X_D).
\ee
This is a relation between two rational functions of $X^Q$ (we remind that $X_D=X^{Q/P}$). Often, they are Laurent polynomials of $X^Q$ which are of the same order, and equating each order gives algebraic relations between $\kappa, \bxi$ and the duals $\tilde \kappa,\tilde \bxi$. More generally, these relations can always be extracted even if we have rationals instead of Laurent polynomials. 
These relations are essentially the same as the ones presented in \cite{hkt,hsx} for local ${\mathbb P}^1 \times {\mathbb P}^1$ and local ${\mathcal B}_3$. Here we put their procedure in a more general context. 
If this does not give enough conditions as in the case of full ${\mathcal B}_3$, one can use in addition the condition of equating the determinant too, or equivalently, the condition given by
\be
	\Delta(X) = \Delta_D(X_D).
\ee
So in the end, even that case can be dealt with using $2 \times 2$ matrices instead of the larger ones given in \cite{hsx}.
In any case, the relations between $\tilde \kappa, \tilde \bxi$ and $ \kappa,  \bxi$ are fully determined by (\ref{trpmDelta}).
Let us give some examples. 

For local ${\mathbb P}^2$ we only have $\kappa$, and no $\bxi$. We find for $a(X)$ and $b(X)$ in (\ref{VVab}):
\be
	a(X)=1+\frac{\kappa}{X}, \qquad b(X)=\frac{q^{-3/2}}{X^3}.
\ee
From this, we can build the matrix ${\mathcal M}(X)$ and its dual, compute their traces and equate them.
The relations in some rational cases are
\be
\ba
	\, & P=1, \, Q=1, \qquad \tilde \kappa =\kappa, \\
	\, & P=1, \, Q=2, \qquad \tilde \kappa =-\kappa^2, \\
	\, & P=1, \, Q=3, \qquad \tilde \kappa =\kappa^3+3, \\
	\, & P=2, \, Q=3, \qquad -\tilde \kappa^2 =\kappa^3-3, \\
	\, & P=1, \, Q=4, \qquad \tilde \kappa =-\kappa^4-4\sqrt{2}\kappa, \\
	\, & P=3, \, Q=4, \qquad \tilde \kappa^3-3 =-\kappa^4+4\sqrt{2}\kappa, \\
\ea
\ee

For local ${\mathbb P}^1 \times {\mathbb P}^1$ we have $\kappa$, and a mass parameter $m \equiv \xi_1$. We find
\be
	a(X)=1+\frac{\kappa}{X}+\frac{m}{X^2}, \qquad b(X)=\frac{q^{-1}}{X^2}.
\ee
The relations in some rational cases are
\be
\ba
	\, & P=1, \, Q=1, \qquad \tilde \kappa =\kappa, \\
	 & \qquad \qquad \qquad \qquad  \tilde m = m, \\
	 \, & P=1, \, Q=2, \qquad \tilde \kappa =-\kappa^2+2(m+1), \\
	 & \qquad \qquad \qquad \qquad  \tilde m = m^2, \\
	  \, & P=1, \, Q=3, \qquad \tilde \kappa =\kappa^3-3(m+1)\kappa, \\
	 & \qquad \qquad \qquad \qquad  \tilde m = m^3, \\
	  \, & P=2, \, Q=3, \qquad -\tilde \kappa^2+2(\tilde m+1)=\kappa^3-3(m+1)\kappa, \\
	 & \qquad \qquad \qquad \qquad  \tilde m^2 = m^3, \\
	   \, & P=1, \, Q=4, \qquad \tilde \kappa=-\kappa^4+4(m+1)\kappa^2-2(m^2+1), \\
	 & \qquad \qquad \qquad \qquad  \tilde m = m^4, \\
	    \, & P=3, \, Q=4, \qquad \tilde \kappa^3-3(\tilde m+1)\tilde \kappa=-\kappa^4+4(m+1)\kappa^2-2(m^2+1), \\
	 & \qquad \qquad \qquad \qquad  \tilde m^3 = m^4, \\
\ea
\ee
For the resolved ${\mathbb C}^3/{\mathbb Z}_5$ we have $\kappa$, and another true modulus $\kappa_1 \equiv \xi_1$. We find
\be
	a(X)=1+\frac{\kappa}{X}+\frac{\kappa_1}{X^2}, \qquad b(X)=\frac{q^{-5/2}}{X^5}.
\ee
The relations in some rational cases are
\be
\ba
	\, & P=1, \, Q=1, \qquad \tilde \kappa =\kappa, \\
	 & \qquad \qquad \qquad \qquad  \tilde \kappa_1 = \kappa_1, \\
	 \, & P=1, \, Q=2, \qquad \tilde \kappa =-\kappa^2+2 \kappa_1, \\
	 & \qquad \qquad \qquad \qquad  \tilde \kappa_1 = \kappa_1^2, \\
	  \, & P=1, \, Q=3, \qquad \tilde \kappa =\kappa^3-3 \kappa \kappa_1, \\
	 & \qquad \qquad \qquad \qquad  \tilde \kappa_1 = \kappa_1^3+3 \kappa, \\
	   \, & P=2, \, Q=3, \qquad -\tilde \kappa^2+2 \tilde \kappa_1 =\kappa^3-3 \kappa \kappa_1, \\
	 & \qquad \qquad \qquad \qquad  \tilde \kappa_1^2 = \kappa_1^3-3 \kappa, \\
	   \, & P=1, \, Q=4, \qquad \tilde \kappa =-\kappa^4+4 \kappa^2 \kappa_1-2\kappa_1^2, \\
	 & \qquad \qquad \qquad \qquad  \tilde \kappa_1 = \kappa_1^4+4\sqrt{2} \kappa \kappa_1, \\
	  \, & P=3, \, Q=4, \qquad \tilde \kappa^3-3 \tilde \kappa \tilde \kappa_1 =-\kappa^4+4 \kappa^2 \kappa_1-2\kappa_1^2, \\
	 & \qquad \qquad \qquad \qquad  \tilde \kappa_1^3-3\tilde \kappa = \kappa_1^4-4\sqrt{2} \kappa \kappa_1, \\
\ea
\ee
We see that here, in contrast with local ${\mathbb P}^1 \times {\mathbb P}^1$, the extra parameter $\kappa_1$ has a non-trivial relation with its dual. This is certainly because it is a true modulus, whereas $m$ in local ${\mathbb P}^1 \times {\mathbb P}^1$ is a simple mass parameter.

\subsection{The ``fully on-shell" eigenfunctions}

To obtain the ``fully on-shell"  eigenfunction from expression (\ref{logPsiRational}), we need to do two more steps. First, add the polynomial part in $x$, which was truncated in (\ref{truncWKB}). Second, since we are in the hyperelliptic case, we should linearly combine it with the second part of the eigenfunction which corresponds to the second solution of the WKB. 
This consists in taking the second branch of the function $y(x)$ when performing the small $\hbar$ WKB. It is not hard to convince oneself that in expression (\ref{logPsiRational}), this corresponds to evaluate all the ingredients of (\ref{logPsiRational}) on their second branch. 
For the integral expressions, the base point should not be changed (it remains at $\infty$ on the first sheet), but the path of integration should extend to the point $\bar X$ on the second sheet, which is the image of $X$ under the obvious involution that exchanges the two sheets of the double cover. This is exactly the prescription which is used in \cite{mz1, mz2} to build eigenfunctions from open topological string data.\footnote{In those references, the reasoning behind this prescription is a priori different: the sum of the two related functions comes from the contribution of two distinct saddles in a certain integral transform.}
Let us denote $\Psi(X)$ the exponential of expression (\ref{logWKB2}). In the rational case, it is the exponential of expression (\ref{logPsiRational}). We propose that the exact eigenfunction is given by
\be
	\label{finalPsi}
	\psi(X) = \re^{(-\ri \hbar)^{-1}s_0(x)+s_1(x)} \left ( \Psi(X)+\Psi(\bar X) \right ).
\ee 
In the rational case, this is a completely explicit expression.

 The point $\bar X$ can be reached through different inequivalent paths when evaluating integrated expressions. Requiring single valuedness of the resulting eigenfunction, we should impose that the difference between two inequivalent integrations give $2\pi \ri \times \rm{integer}$. In this way, the final eigenfunction will not depend on the path chosen to reach $\bar X$. This leads to the well known argument of monodromy invariance, and yields quantization conditions for all the true moduli.

Let us now proceed to the testing of this construction in some examples. The eigenfunctions and eigenvalues are compared to purely numerical results which can be obtained using the hamiltonian truncation method in the basis of the harmonic oscillator (appropriately scaled Hermite functions). This method is explained for example in \cite{hw,KP}, where it is used for the same kind of difference equations as the ones considered here.
 
 
 \sectiono{Examples}
 \label{sect4}
 
 \subsection{Local ${\mathbb P}^2$}
 
 The difference equation related to the geometry called local $\mathbb P^2$ is the simplest example of our family of difference equations, since it has only one modulus $\kappa$.
 So the ``fully on-shell" eigenfunctions are exactly the on-shell eigenfunctions.
 In an appropriate parametrization, its mirror curve is given by the zero locus of
 \be
 	W(x,y) =\re^{x}+\re^{y}+\re^{-x-y}+\kappa,
 \ee
 which, after quantization, leads to the difference equation
  \be
 	\re^{x} \psi(x)+\psi(x-\ri \hbar)+\re^{-\frac{\ri \hbar}{2}} \re^{-x} \psi(x+\ri \hbar) =-\kappa \psi(x).
 \ee
 If we look at this system as the quantization of some classical one dimensional system, then classically, the region allowed in the real phase space $(x,y)$ is non empty  for $\kappa<-3$. We will assume this regime for $\kappa$.
 We now build the exact eigenfunctions and quantization conditions for the spectrum using the technology developed in the previous sections.
 \newline
 
 As a warm up, let us consider the case $\hbar=2\pi$. This is the self dual case, where $P=Q=1$. So every quantity is identified with its dual.
After computing every constituent of (\ref{logPsiRational}), we obtain
\be
	\label{logPsiP2h2pi}
	\log \Psi(X) = \frac{\ri}{2\pi}\left ( \int_\infty^X \log(X') \left ( -\frac{3}{2X'}+\frac{3X'+\kappa}{2\sqrt{\sigma(X')}}\right )\rd X' +\lambda \int_{\infty}^X \frac{\rd X'}{\sqrt{\sigma(X')}}\right )+\frac{1}{2}\log \left ( \frac{X^4}{\sigma(X)}\right),
\ee
where
\be
	\sigma(X)=4X+X^2(X+\kappa)^2 \equiv X\prod_{n=1}^3(X-A_n).
\ee
For the regime of $\kappa$ we are interested in, we have $A_1<0$ and $\bar A_3=A_2$, with positive real part. We take the $\mathcal A$ cycle to be the one which encircles $A_2$ and $A_3$ counterclockwise, and the $\mathcal B$ the one which encircles $A_1$ and $A_2$. We define the following $A$ and $B$ periods
\be
\ba
	\Pi_{A,B} &=\oint_{\mathcal A,\mathcal B} \log(X) \left ( -\frac{3}{2X}+\frac{3X+\kappa}{2\sqrt{\Delta(X)}}\right )\rd X, \\
	\Pi^{(\lambda)}_{A,B} &=\oint_{\mathcal A,\mathcal B} \frac{\rd X}{\sqrt{\Delta(X)}}.
\ea
\ee
The last term in (\ref{logPsiP2h2pi}) with the logarithm function does not contribute to monodromy.\footnote{More precisely, it contributes with integer multiples of $2\pi \ri$ which are trivial.} 
Monodromy invariance is expressed as
\be
	\label{Phimonodr}
\ba
	\log \Psi_{\rm WKB}{\Large |_{\mathcal A}} =\frac{\ri}{2\pi} \left ( \Pi_{A}+\lambda \Pi^{(\lambda)}_{A} \right )= 2 \pi \ri M ,\\
	\log \Psi_{\rm WKB}{\Large |_{\mathcal B}} =\frac{\ri}{2\pi} \left ( \Pi_{B}+\lambda \Pi^{(\lambda)}_{B}  \right )= 2 \pi \ri N,
\ea
\ee
where $M,N$ are integers. It turns out that the $A$ periods are purely imaginary, so this sets $M$=0. This gives an equation to fix the value of $\lambda$. Also, it turns out that the combination of $B$ periods in the second line above is positive. So the quantization condition is
\be
	\frac{\ri}{2\pi} \left ( \Pi_{B}+\left (-\frac{\Pi_{A}}{\Pi^{(\lambda)}_{A}} \right ) \Pi^{(\lambda)}_{B}  \right )= 2 \pi \ri (n+1).
\ee
For each non negative integers $n=0,1,2,\ldots$, this equation fixes a value for $\kappa$ which we call $\kappa_n$. It can be checked numerically that this corresponds to the eigenvalues of the spectral problem. Here are some values found by performing the integration numerically and solving the equation using Newton's method:
\be
\ba
	E_0 &=\log(-\kappa_0) = 2.56264206862381937081\ldots, \\
	E_1 &=\log(-\kappa_1) = 3.91821318829983977872\ldots, \\
	E_2 &=\log(-\kappa_2) = 4.91178982376733605820\ldots, \\
	E_3 &=\log(-\kappa_3) = 5.73573703542155946556\ldots, \\
	\ldots
\ea
\ee
These values, and the quantization condition itself, agree with the literature, See for example \cite{ghm}, where the periods are written down explicitly using hypergeometric and Meijer functions. The eigenfunctions themselves can be seen to agree with the on-shell results in \cite{mz2} up to some overall phase.
\newline

Let us now consider a more involved case: $\hbar=3\pi/2$. This is a non-trivial case, since both $P$ and $Q$ are different from $1$, namely $P=3$ and $Q=4$. It is dual to the case $\hbar=8\pi/3$. The relation between $\kappa$ and its dual can be found in section \ref{sectRels}. Let us define
\be
	P_{3,4}(\kappa) = -\kappa^4+4\sqrt{2}\kappa, \qquad \qquad
	P_{4,3}(\tilde \kappa) = \tilde \kappa^3-3,
\ee
 so that the relation is
 \be
 	P_{3,4}(\kappa)=P_{4,3}(\tilde \kappa).
\ee
We also have $X_D=X^{4/3}$.
 Let us  define
 \be
 	\label{sigmaP2h3po2}
 \ba
 	\sigma(X) &= -4 X^4 +X^8 (X^4+P_{3,4}(\kappa))^2 \\
		& \equiv X^4 \prod_{k=0}^3 \prod_{n=1}^3(X-\re^{\frac{\pi \ri}{2}k}A_n),
\ea
 \ee
 and
 \be
 \ba
 	p_1(X) &=\kappa X^{10}-\kappa^2 X^9 +(6 \sqrt{2}\kappa^2-\kappa^5)X^6+(-3-6\sqrt{2}\kappa^3+\kappa^6)X^5 \\
	& \quad +15 \kappa X^4-5 \kappa^2 X^3+(-3\sqrt{2}+\kappa^3)X^2-\sqrt{2}\kappa X, \\
	p_2(X) &= X^4-\kappa X^3+\kappa^2 X^2+(\sqrt{2}-\kappa^3)X-\sqrt{2}\kappa, \\
	\tilde p_1(X_D) &= \tilde \kappa X^7_D +(\tilde \kappa^4-8 \tilde \kappa)X_D^4+3\tilde \kappa^2 X_D^3-\tilde \kappa^3 X_D^2-2 \tilde \kappa X_D,\\
	\tilde p_2(X_D) &= X_D^3 - \tilde \kappa X_D^2+ \tilde \kappa^2 X_D-1. \\
\ea
 \ee
 Using the formulas given in the previous sections, we find
 \be
 \ba
 	\varphi_Q(X) &= \frac{1}{4} \log \left [ \frac{1}{2}\left ( 1+ \frac{P_{3,4}(\kappa)}{X^4} \right ) +\frac{1}{2X^8} \sqrt{\Delta(X)} \right ], 
 \ea
 \ee
 as well as, after some tedious work of simplifications,
  \be
 \ba
	 -\frac{1}{2} \varphi_Q(X)
		 +& \int_{\infty}^{X} \ri \partial_\hbar \varphi_Q(X') \frac{\rd X'}{X'}
		+ \sum_{k=1}^{Q-1} \frac{\varphi_k(X)}{q^k-1} \\
		& = \frac{1}{4} \log \left (\frac{X^8 p_2(X)}{\Delta(X)} \right ) 
		+\int_\infty^X \frac{-\ri}{2}\frac{p_1(X')}{p_2(X')\sqrt{\sigma(X')}}\rd X', \\
		 -\frac{P}{2Q} \varphi_Q(X)
		 +& \int_{\infty}^{X_D} \ri \partial_{\hbar_D} \tilde \varphi_P(X_D') \frac{\rd X_D'}{X_D'} + \sum_{k=1}^{P-1} \frac{\tilde \varphi_k(X_D)}{{q_D}^k-1} \\
		 &=  \frac{1}{4} \log \left (\frac{X_D^6 \tilde p_2(X_D)}{\sigma(X)} \right )
		   +  \int_\infty^{X_D} \frac{\ri}{2\sqrt{3}}\frac{ \tilde p_1(X_D')}{\tilde p_2(X_D')\sqrt{\sigma({X_D'}^{3/4})}}\rd X_D' .
 \ea
 \ee
Putting everything together, we obtain:
\be
	\label{p2logpsi3o2}
\ba
	\log \Psi(X) &= \frac{2\ri}{3\pi} \left ( \int_\infty^X \log(X') \left ( -\frac{3}{2X'}+\frac{{X'}^3(3{X'}^4+P_{3,4}(\kappa))}{2\sqrt{\sigma(X')}}\right )\rd X' +\lambda \int_{\infty}^X \frac{\frac{1}{4} P_{3,4}'(\kappa){X'}^3 \rd X'}{\sqrt{\sigma(X')}}\right ) \\
	& \qquad +  \int_\infty^X \frac{-\ri}{2}\frac{p_1(X')}{p_2(X')\sqrt{\sigma(X')}}\rd X' +  \int_\infty^X \frac{2\ri}{3\sqrt{3}}\frac{\tilde p_1({X'}^{4/3}) {X'}^{1/3}}{\tilde p_2({X'}^{4/3})\sqrt{\sigma(X')}}\rd X'   \\
	& \qquad + \frac{1}{2} \log \left ( \frac{X^8 p_2(X) \tilde p_2(X^{4/3})}{\sigma(X)} \right ).
\ea
\ee
In the regime of $\kappa$ we are interested in, the $A_n$ defined in (\ref{sigmaP2h3po2}) are ordered as $0<A_1<A_2<A_3$. We take the $\mathcal A$ cycle to be the one which encircles $A_2$ and $A_3$ counterclockwise, and the $\mathcal B$ cycle the one which encircles $A_1$ and $A_2$. We define the following $A$ and $B$ periods
\be
\ba
	\Pi_{A,B} &=\oint_{\mathcal A,\mathcal B} \left [ \log(X) \left ( -\frac{3}{2X}+\frac{{X}^3(3{X}^4+P_{3,4}(\kappa))}{2\sqrt{\sigma(X)}} \right ) \right . \\
	& \qquad  \qquad \qquad \left.+ \frac{-3\pi}{4}\frac{p_1(X)}{p_2(X)\sqrt{\sigma(X)}}+\frac{\pi}{\sqrt{3}}\frac{\tilde p_1({X}^{4/3}) {X}^{1/3}}{\tilde p_2({X}^{4/3})\sqrt{\sigma(X)}} \right  ] \rd X, \\
	\Pi^{(\lambda)}_{A,B} &=\oint_{\mathcal A,\mathcal B} \frac{\frac{1}{4} P_{3,4}'(\kappa){X}^3 \rd X}{\sqrt{\sigma(X)}}.
\ea
\ee
With these definitions, monodromy invariance is given by eq. (\ref{Phimonodr}) with a factor $4/3$:
\be
\ba
	\log \Psi_{\rm WKB}{\Large |_{\mathcal A}} =\frac{2\ri}{3\pi} \left ( \Pi_{A}+\lambda \Pi^{(\lambda)}_{A} \right )= 2 \pi \ri M ,\\
	\log \Psi_{\rm WKB}{\Large |_{\mathcal B}} =\frac{2\ri}{3\pi} \left ( \Pi_{B}+\lambda \Pi^{(\lambda)}_{B}  \right )= 2 \pi \ri N,
\ea
\ee
Empirically, the $A$ periods are always imaginary in our regime of $\kappa$, so we get the following quantization condition
\be
	\frac{2\ri}{3\pi} \left ( \Pi_{B}+\left (-\frac{\Pi_{A}}{\Pi^{(\lambda)}_{A}} \right ) \Pi^{(\lambda)}_{B}  \right )= 2 \pi \ri n,
\ee
for $n=0,1,2,\ldots$. This equation fixes a value for $\kappa$ which we call $\kappa_n$. Here are some values found by numerical integration and solving the equation using Newton's method:
\be
	\label{P2kappa3pio2}
\ba
	E_0 &=\log(-\kappa_0) = 2.23447824285951068410\ldots, \\
	E_1 &=\log(-\kappa_1) = 3.40332799272918290269\ldots, \\
	E_2 &=\log(-\kappa_2) = 4.26178057406546295246\ldots, \\
	\ldots
\ea
\ee
\begin{figure}[h]
\begin{center}
\includegraphics[scale=1]{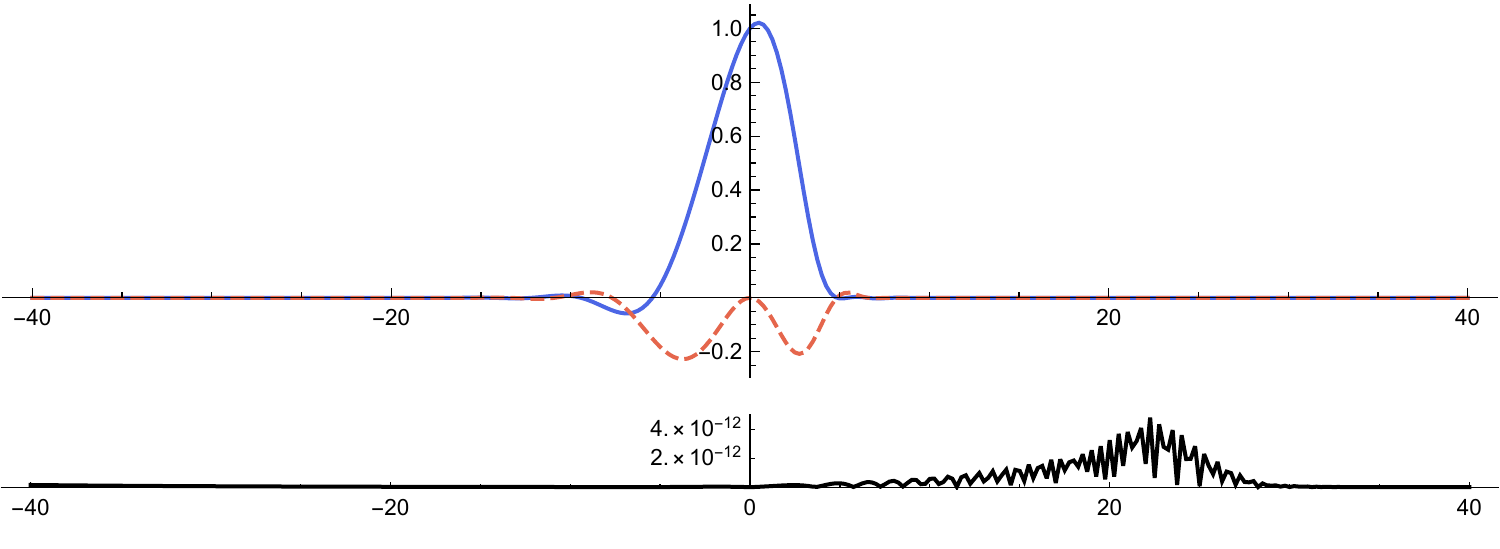}
\caption{Local $\mathbb P^2$ at $\hbar=3\pi/2$: exact ground state eigenfunction (real part in blue and imaginary part in red) and the absolute difference with numerics coming from numerical diagonalization of a $250 \times 250$ matrix (rescaled to match the exact eigenfunctions at $x=0$). For this size of the matrix, the maximal difference is of the order $10^{-12}$. }
\label{figP2n0}
\end{center}
\end{figure}
\begin{figure}[h]
\begin{center}
\includegraphics[scale=1]{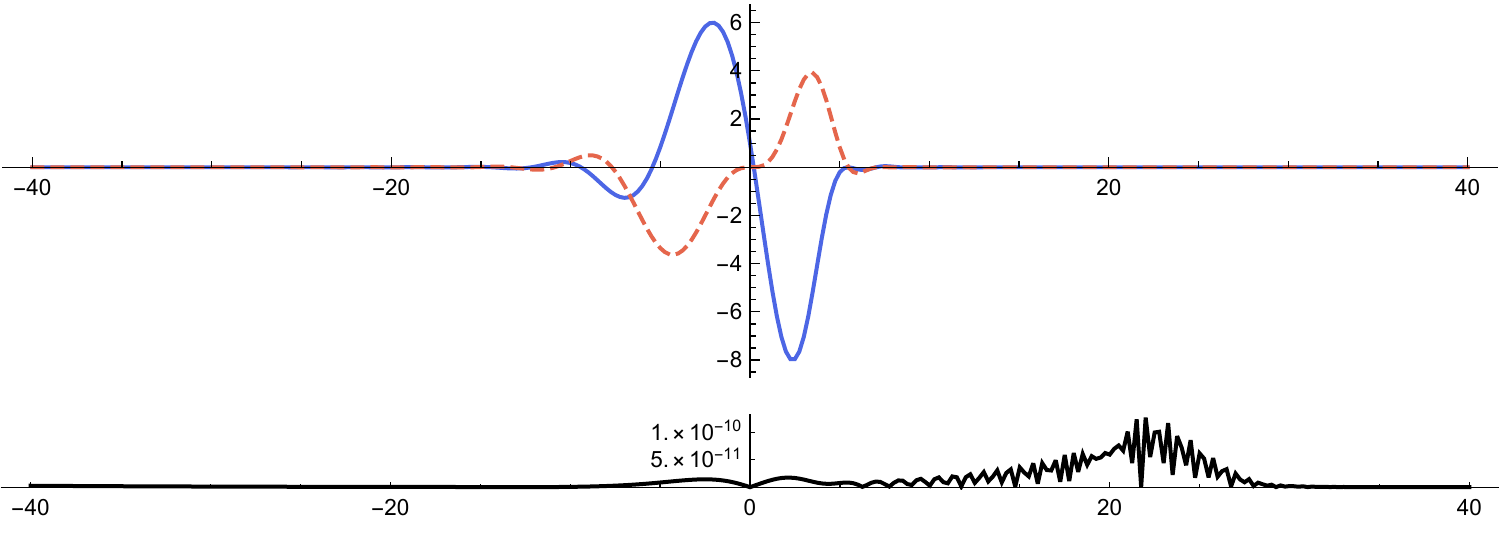}
\caption{Local $\mathbb P^2$ at $\hbar=3\pi/2$: same plot as in Fig. 1, for the first excited state. The maximal difference is of the order $10^{-10}$.  }
\label{figP2n1}
\end{center}
\end{figure}

\noindent
These have been checked using numerical hamiltonian truncation. We can also test the eigenfunction given by formula (\ref{finalPsi}). The symmetric sum in (\ref{finalPsi}) as well as monodromy invariance ensure that the final eigenfunction is free of branch-points, single valued and analytic on the $X$ plane (at least in a sector of the $X$ plane containing the positive real line). The exact eigenfunctions can be seen in Fig. \ref{figP2n0} and Fig.  \ref{figP2n1}. They are obtained by evaluating numerically the integral expressions in (\ref{p2logpsi3o2}). The proposed exact expression reproduces well the numerical result obtained using numerical hamiltonian truncation. The difference between the exact results and the purely numerical results decrease as we increase the size of the numerical truncated hamiltonian matrix.

 \subsection{Local ${\mathbb P}^1 \times {\mathbb P}^1$}
 
The difference equation related to local $\mathbb P^1 \times \mathbb P^1$ is also a simple example, since we do not have any additional true moduli, only a mass parameter $\xi_1=m$. The ``fully on-shell" eigenfunctions are exactly the on-shell eigenfunctions. We assume that $m$ is positive. Also, this case is related to the relativistic Toda lattice with two particles. 
In an appropriate parametrization, the mirror curve of local $\mathbb P^1 \times \mathbb P^1$ is given by the zero locus of
\be
 	W(x,y) =\re^{x}+m \re^{-x}+\re^{y}+\re^{-y}+\kappa,
 \ee
 which leads to the difference equation
  \be
 	(\re^{x} +m \re^{-x})\psi(x)+\psi(x-\ri \hbar)+\psi(x+\ri \hbar) =-\kappa \psi(x).
 \ee
 If we look at this system as the quantization of some classical one dimensional system, then classically, the region allowed in the real phase space $(x,y)$ is non-empty  for $\kappa<-2-2\sqrt{m}$. We will assume this regime for $\kappa$.
 We now build the exact eigenfunctions and quantization conditions for the spectrum using the technology developed in the previous sections.
\newline

Let us again work out the self dual case $\hbar=2\pi$. We obtain
\be
\ba
	\log \Psi(X) &= \frac{\ri}{2\pi}\left ( \int_\infty^X \log(X') \left ( -\frac{1}{X'}+\frac{{X'}^2-m}{X'\sqrt{\sigma(X')}}\right )\rd X' \right. \\
	& \qquad \left.  +\lambda \int_{\infty}^X \frac{\rd X'}{\sqrt{\sigma(X')}}+\lambda_m \int_{\infty}^X \frac{\rd X'}{X'\sqrt{\sigma(X')}}\right ) \\
	& \qquad +\frac{1}{2}\log \left ( \frac{X^4}{\sigma(X)}\right),
\ea
\ee
where
\be
	\sigma(X)=-4X^2+(m+X(X+\kappa))^2=\prod_{n=1}^4(X-A_n).
\ee
In the regime of $\kappa$ we are interested in, all the branch-points $A_n$ are positive real. We order them increasingly. We define the cycle $\mathcal A$ encircling $A_3$ and $A_4$ counterclockwise, the cycle $\tilde {\mathcal A}$ encircling $A_1$ and $A_2$ counterclockwise, and the cycle $\mathcal B$ encircling $A_2$ and $A_3$. Indeed, since we have two undetermined constants $\lambda$ and $\lambda_m$, we need three monodromy conditions for the quantization condition. Fortunately, the cycles $\mathcal A$ and $\tilde {\mathcal A}$ are inequivalent since the integral in front of $\lambda_m$ picks up a residue at the pole $X=0$ when we deform the cycle $\mathcal A$ to $\tilde {\mathcal A}$. It is then better to consider the set $\mathcal A$, ${\mathcal A} + \tilde {\mathcal A}$ and ${\mathcal B}$ as the set of independent cycles.  Let us define
\be
	\label{localP1periods2pi}
\ba
	\Pi_{A,A+{\tilde A},B} &=\oint_{\mathcal A,{\mathcal A}+{\tilde {\mathcal A}},\mathcal B} \log(X) \left ( -\frac{1}{X}+\frac{{X}^2-m}{X\sqrt{\sigma(X)}}\right )\rd X, \\
	\Pi^{(\lambda)}_{A,A+{\tilde A},B} &=\oint_{\mathcal A,{\mathcal A}+{\tilde {\mathcal A}},\mathcal B} \frac{\rd X}{\sqrt{\sigma(X)}}, \\
	\Pi^{(\lambda_m)}_{A,A+{\tilde A},B} &=\oint_{\mathcal A,{\mathcal A}+{\tilde {\mathcal A}},\mathcal B} \frac{\rd X}{X\sqrt{\sigma(X)}}.
\ea
\ee
Monodromy invariance is expressed as
\be
	\label{PhimonodrP1}
\ba
	\log \Psi_{\rm WKB}{\Large |_{\mathcal A}} &=\frac{\ri}{2\pi} \left ( \Pi_{A}+\lambda \Pi^{(\lambda)}_{A} + \lambda_m \Pi_{A}^{(\lambda_m)}\right )= 2 \pi \ri M ,\\
		\log \Psi_{\rm WKB}{\Large |_{{\mathcal A}+\tilde {\mathcal A}}} &=\frac{\ri}{2\pi} \left ( \Pi_{A+\tilde A}+\lambda \Pi^{(\lambda)}_{A+\tilde A} +\lambda_m \Pi_{A+\tilde A}^{(\lambda_m)}\right )= 2 \pi \ri {\tilde M} ,\\
	\log \Psi_{\rm WKB}{\Large |_{\mathcal B}} &=\frac{\ri}{2\pi} \left ( \Pi_{B}+\lambda \Pi^{(\lambda)}_{B}  +\lambda_m \Pi_{B}^{(\lambda_m)} \right )= 2 \pi \ri N,
\ea
\ee
where $M,\tilde M,N$ are integers. It turns out that the $A$ and $\tilde A$ periods are purely imaginary whereas $\lambda$ and $\lambda_m$ should be real, so this fixes $M$=$\tilde M$=0. 
It is actually easy to compute the periods for ${\mathcal A} + \tilde {\mathcal A}$. By deforming the contour and taking the residue at $X=0$, we find
\be
	\Pi^{(\lambda)}_{A+{\tilde A}} = 0, \qquad \qquad \Pi^{(\lambda_m)}_{A+{\tilde A}} = \frac{2\pi \ri}{m}.
\ee
Also, using that $\sigma(m/X)=m^2  \sigma(X)/X^4$, we find after a change of variables
\be
	\Pi_{A+{\tilde A}} = -2\pi \ri \log m.
\ee
Then, the quantization condition for the cycle $\mathcal A+\tilde{\mathcal A}$ yields
\be
	\lambda_m = m \log m.
\ee
It turns out that the combination of $B$ periods in the third line of (\ref{PhimonodrP1}) is positive for our regime of $\kappa$. So the remaining two monodromy conditions give the following quantization condition:
\be
	\label{simpQuantpi}
	\frac{\ri}{2\pi} \left (  \Pi_{B}-\left (\frac{\Pi_A+m \log (m) \Pi_A^{(\lambda_m)}}{\Pi_{A}^{(\lambda)}} \right )\Pi^{(\lambda)}_{B}
	+m \log(m) \Pi_{B}^{(\lambda_m)} \right ) = 2\pi \ri (n+1).
\ee
In the case $m=1$ we retrieve the results of \cite{ks}. By numerical computation of the periods and Newton's method for solving the quantization condition, we can get the eigenvalues $\kappa_n$, \mbox{$n=0,1,2,\ldots$}. Here we list some results, which have been checked using numerical diagonalization.
For $m=1$,
\be
\ba
	E_0 &=\log(-\kappa_0) = 2.88181542992629678247\ldots, \\
	E_1 &=\log(-\kappa_1) = 4.25459152858199378358\ldots, \\
	E_2 &=\log(-\kappa_2) = 5.28819530714418547625\ldots, \\
	\ldots
\ea
\ee
For $m=1/3$,
\be
\ba
	E_0 &=\log(-\kappa_0) = 2.62164098025513043508\ldots, \\
	E_1 &=\log(-\kappa_1) = 3.98889597312465176636\ldots, \\
	E_2 &=\log(-\kappa_2) = 5.02068317784445369322\ldots, \\
	\ldots
\ea
\ee
The eigenfunctions match the numerical results obtained by hamiltonian truncation. For the case $m=1$, up to an overall phase, the eigenfunction is equivalent to the exact result of \cite{ks} and to the on-shell restriction of the result in \cite{mz1}.
\newline 

Let us now take $\hbar=2\pi/3$, which is a more involved case. We have $P=1$ and $Q=3$. Also, $\tilde \kappa=\kappa^3-3(m+1)\kappa$, $\tilde m=m^3$ and $X_D=X^3$. Let us define
\be
\ba
	\sigma(X) &=-4X^6+(X^6+X^3 (\kappa^3-3(m+1)\kappa)+m^3)^2 \\
	& \equiv \prod_{n=1}^4(X-A_n)(X-\re^{\frac{2\pi \ri}{3}}A_n)(X-\re^{\frac{4\pi \ri}{3}}A_n),
\ea
\ee
and
\be
\ba
	p_1(X) &= X^8 \kappa+ X^7(2-2m)+X^6 m \kappa+X^5(\kappa^4-(5+3m)\kappa^2+4-4m) \\
		& \qquad +X^4(-2m \kappa^3+6 m^2 \kappa+14 m \kappa)+X^3(m\kappa^4-(5+3m)m\kappa^2+4m -4m^2) \\
		& \qquad +X^2 m^3 \kappa+X m^3(2-2m)+m^4 \kappa, \\
	p_2(X) &=X^4-X^3 \kappa+X^2(\kappa^2-m-1)-X \kappa m+m^2.
\ea
\ee
Using (\ref{logPsiRational}) and the methods given in the previous sections to compute the different ingredients, we obtain after some simplifications:
\be
\ba
	\log \Psi(X) &= \frac{3\ri}{2\pi}\left ( \int_\infty^X \log(X') \left ( -\frac{1}{X'}+\frac{{X'}^6-m^3}{X'\sqrt{\sigma(X')}}\right )\rd X' \right. \\
	& \qquad \left.  + \lambda \int_{\infty}^X \frac{(\kappa^2-m-1) {X'}^2  \rd X'}{\sqrt{\sigma(X')}}+\lambda_m \int_{\infty}^X \frac{(m^2-{X'}^3 \kappa)\rd X'}{X'\sqrt{\sigma(X')}}\right ) \\
	& \qquad + \int_{\infty}^X \frac{\ri}{2 \sqrt{3}} \frac{p_1(X)}{p_2(X)\sqrt{\sigma(X)}} \rd X'
	+\frac{1}{2}\log \left ( \frac{X^8 p_2(X)}{\sigma(X)}\right).
\ea
\ee
In a suitable regime, we have real positive $A_n$ which we order increasingly. As before, we define the cycle $\mathcal A$ encircling $A_3$ and $A_4$ counterclockwise, the cycle $\tilde {\mathcal A}$ encircling $A_1$ and $A_2$ counterclockwise, and the cycle $\mathcal B$ encircling $A_2$ and $A_3$.
Here we have an additional contribution to monodromy given by the first term of the last line. We define
\be
	\label{periods2pio3}
\ba
	\Pi_{A,{\tilde A},B} &=\oint_{{\mathcal A}, {\tilde {\mathcal  A}},{\mathcal B}} \left [ \log(X) \left ( -\frac{1}{X}+\frac{{X}^6-m^3}{X\sqrt{\sigma(X)}}\right ) + \frac{\pi}{3 \sqrt{3}} \frac{p_1(X)}{p_2(X)\sqrt{\sigma(X)}}  \right ] \rd X , \\
	\Pi^{(\lambda)}_{A,{\tilde A},B} &=\oint_{\mathcal A,{\tilde {\mathcal A}},\mathcal B}  \frac{(\kappa^2-m-1) {X}^2  \rd X}{\sqrt{\sigma(X)}}, \\
	\Pi^{(\lambda_m)}_{A,{\tilde A},B} &=\oint_{\mathcal A,{\tilde {\mathcal A}},\mathcal B}  \frac{(m^2-{X}^3 \kappa)\rd X}{X\sqrt{\sigma(X)}}.
\ea
\ee
\begin{figure}[h]
\begin{center}
\includegraphics[scale=1]{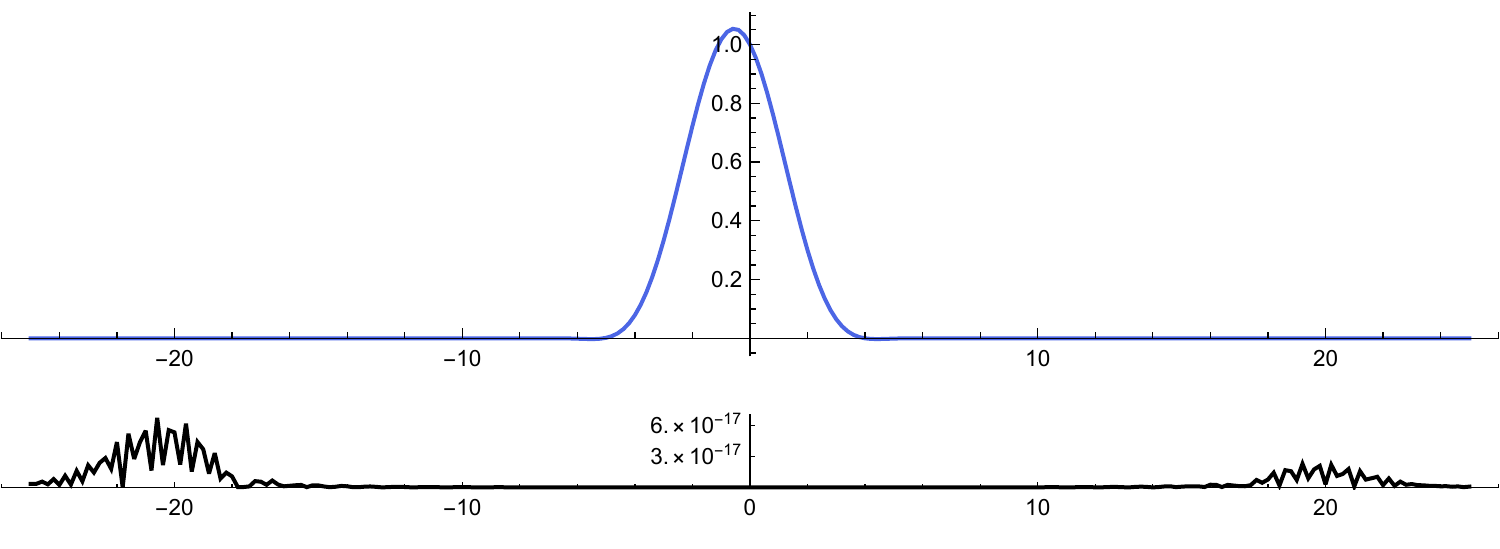}
\caption{Local $\mathbb P^1 \times \mathbb P^1 $ at $\hbar=2\pi/3$ and $m=1/3$: exact ground state eigenfunction (which is purely real) and the absolute difference with numerics coming from numerical diagonalization of a $200 \times 200$ matrix (rescaled to match the exact eigenfunctions at $x=0$). For this size of the matrix, the maximal difference is of the order $10^{-17}$. }
\label{figP1xP1n0}
\end{center}
\end{figure}
\begin{figure}[h]
\begin{center}
\includegraphics[scale=1]{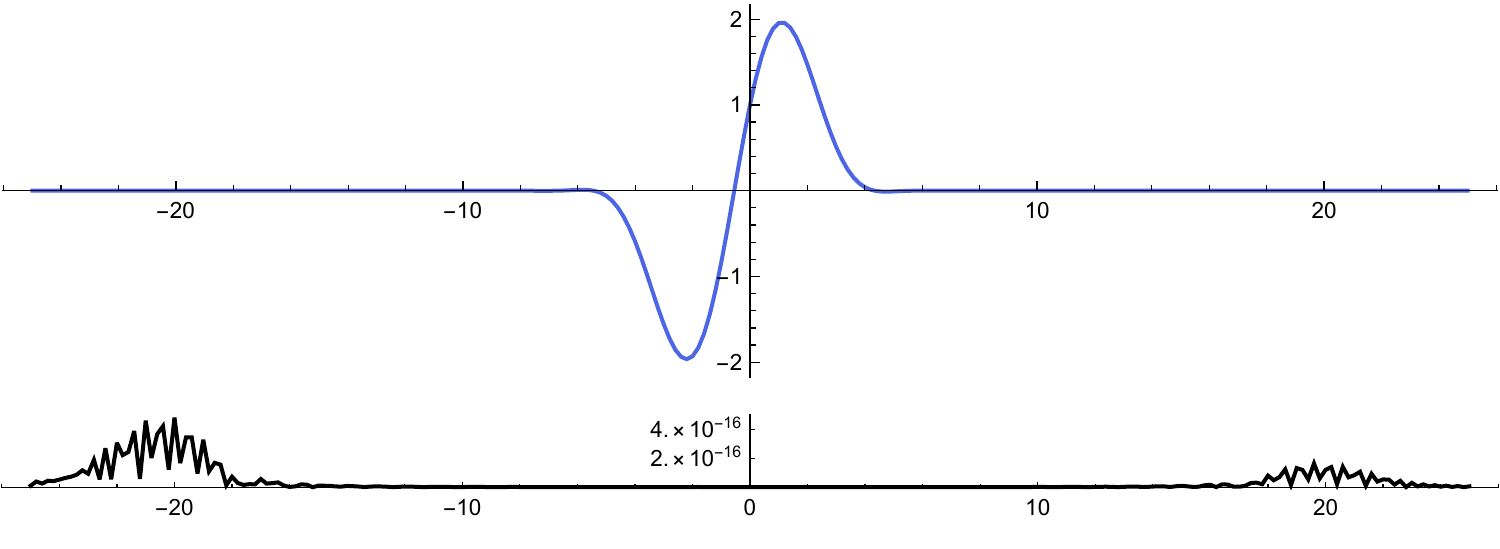}
\caption{Local $\mathbb P^1 \times \mathbb P^1 $ at $\hbar=2\pi/3$ and $m=1/3$: same plot as in Fig. 3, for the first excited state. The maximal difference is of the order $10^{-16}$.  }
\label{figP1xP1n1}
\end{center}
\end{figure}

\begin{figure}[h]
\begin{center}
\includegraphics[scale=0.75]{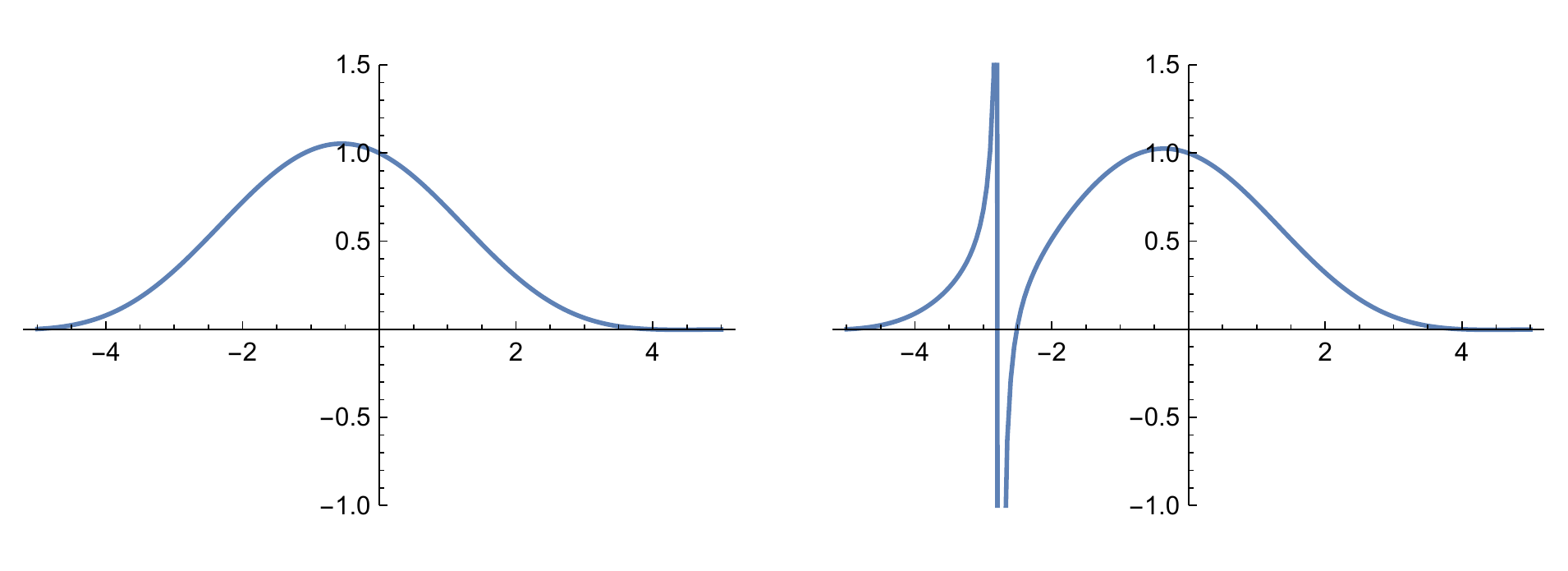}
\caption{Local $\mathbb P^1 \times \mathbb P^1$ at $\hbar=2\pi/3$ and $m=1/3$: On the left, the eigenfunction computed using $\kappa=\kappa_0 \approx -5.22487$ (on-shell value), on the right, the same function computed for a generic value $\kappa=-5.5$ (off-shell value). Our expression for the eigenfunction is singular when off-shell.}
\label{onoffshellPlot}
\end{center}
\end{figure}

\noindent
Monodromy invariance is expressed as
\be
\ba
	\log \Psi_{\rm WKB}{\Large |_{\mathcal A}} =\frac{3\ri}{2\pi} \left ( \Pi_{A}+\lambda \Pi^{(\lambda)}_{A} + \lambda_m \Pi_{A}^{(\lambda_m)}\right )= 2 \pi \ri M ,\\
		\log \Psi_{\rm WKB}{\Large |_{\tilde {\mathcal A}}} =\frac{3\ri}{2\pi} \left ( \Pi_{\tilde A}+\lambda \Pi^{(\lambda)}_{\tilde A} +\lambda_m \Pi_{\tilde A}^{(\lambda_m)}\right )= 2 \pi \ri {\tilde M} ,\\
	\log \Psi_{\rm WKB}{\Large |_{\mathcal B}} =\frac{3\ri}{2\pi} \left ( \Pi_{B}+\lambda \Pi^{(\lambda)}_{B}  +\lambda_m \Pi_{B}^{(\lambda_m)} \right )= 2 \pi \ri N,
\ea
\ee
Again, the $A$ and $\tilde A$ periods are purely imaginary, so $M=\tilde M=0$ and the first two equations determine $\lambda, \lambda_{m}$. The last line gives the quantization condition, which again can be written as
\be
\label{quantp1xp1}
	\frac{3\ri}{2\pi} \left (  \Pi_{B}-\left (\frac{\Pi_A+m \log (m) \Pi_A^{(\lambda_m)}}{\Pi_{A}^{(\lambda)}} \right )\Pi^{(\lambda)}_{B}
	+m \log(m) \Pi_{B}^{(\lambda_m)} \right ) = 2\pi \ri (n+1),
\ee
for $n=0,1,2,\ldots$.
Numerical implementation of the integration and then solving the quantization condition yields the spectrum. 
\noindent
Here are some examples, which have been checked in the usual way.
For $m=1$,
\be
\ba
	E_0 &=\log(-\kappa_0) = 1.90354643917859092548\ldots, \\
	E_1 &=\log(-\kappa_1) = 2.61019754103359928676\ldots, \\
	E_2 &=\log(-\kappa_2) = 3.17373350397478965748\ldots, \\
	\ldots
\ea
\ee
For $m=1/3$,
\be
\ba
	E_0 &=\log(-\kappa_0) = 1.653431255487499979601\ldots, \\
	E_1 &=\log(-\kappa_1) = 2.351194617546936444270\ldots, \\
	E_2 &=\log(-\kappa_2) = 2.911361623248592459660\ldots, \\
	\ldots
\ea
\ee
We can also test the eigenfunction given by formula (\ref{finalPsi}). The symmetric sum in (\ref{finalPsi}) as well as monodromy invariance ensure that the final eigenfunction is free of branch-points, single valued and analytic on the $X$ plane (at least in a sector of the $X$ plane containing the positive real line). Examples of exact eigenfunctions can be seen in Fig. \ref{figP1xP1n0} and Fig.  \ref{figP1xP1n1}. The difference between the exact results and the purely numerical results decrease as we increase the size of the numerical truncated hamiltonian. In Fig. \ref{onoffshellPlot}, we show the importance of monodromy invariance in our construction: we compare an eigenfunction which is on-shell against the evaluation of our expression for the eigenfunction for a generic value of $\kappa$. When $\kappa$ is generic, monodromy invariance is not ensured, and our expression develops a singularity. Therefore, it is not a good eigenfunction for the difference equation.
%

 \subsection{Resolved ${\mathbb C}^3 /  {\mathbb Z}_5$}
 
 The resolved ${\mathbb C}^3 / {\mathbb Z}_5$ geometry is the simplest genus $2$ example. 
This is interesting, since we have a second true modulus $\kappa_1$.
Therefore, the ``fully on-shell" eigenfunctions are only a subset of the on-shell eigenfunctions. 
 In an appropriate parametrization, its mirror curve is given by
\be
 	W(x,y) =\re^{x}+\re^{y}+\re^{-3x-y}+\kappa_1 \re^{-x}+\kappa,
 \ee
 which leads to the difference equation
  \be
 	(\re^{x} +\kappa_1 \re^{-x})\psi(x)+\psi(x-\ri \hbar)+\re^{-\frac{3\ri \hbar}{2}}\re^{-3x}\psi(x+\ri \hbar) =-\kappa \psi(x).
 \ee
 Finding the on-shell eigenfunctions corresponds to finding square integrable solutions $\psi(x)$ to this equation. In that case, $\kappa_1$ acts as a parameter, and $-\kappa$ is quantized. Finding the ``fully-on shell" eigenfunctions corresponds to finding a subset of the on-shell eigenfunctions which have enhanced decay. This extra condition also quantizes $\kappa_1$. These ``fully on-shell" eigenfunctions should also satisfy the dual difference equation.
 
 We will now build some ``fully on-shell" eigenfunctions using the formulas of the previous sections. In practice, we will restrict ourselves to the regime where $\kappa$ is negative with large absolute value and $\kappa_1$ is positive, since it was found to be the good regime from the point of view of the underlying cluster integrable system \cite{fhm}. 
\newline

Let us look at the case where $\hbar=\pi$. We have $P=1$ and $Q=2$, so the relations between the parameters are  $\tilde \kappa =-\kappa^2+2 \kappa_1$ and $\tilde \kappa_1 = \kappa_1^2$. We obtain
\be
\ba
	\log \Psi(X) &= \frac{\ri}{\pi} \left ( \int_{\infty}^X \log X' \left (-\frac{5}{2X'}+\frac{X'(5{X'}^4+3 {X'}^2(-\kappa^2+2\kappa_1)+\kappa_1^2}{2\sqrt{\sigma(X')}} \right )  \rd X' \right. \\
	& \qquad \left. +\lambda \int_{\infty}^X \frac{-2 \kappa {X'}^3 \rd X'}{\sqrt{\sigma(X')}}+\lambda_1 \int_{\infty}^X \frac{2 X' ({X'}^2+\kappa_1)\rd X'}{\sqrt{\sigma(X')}} \right  ) \\
	& \qquad + \int_{\infty}^X \frac{\ri}{2} \frac{5{X'}^2-3 {X'}\kappa+\kappa_1}{({X'}^2-X' \kappa +\kappa_1) \sqrt{\sigma(X')}}\rd X'+ \frac{1}{2}\log \left ( \frac{X^{10}(X^2-X \kappa+\kappa_1)}{\sigma(X)} \right ),
\ea
\ee
with
\be
\ba
	\sigma(X) &= -4X^2+X^4(X^4+(-\kappa^2+2\kappa_1) X^2 + \kappa_1^2)^2 \\
	& \equiv X^2 \prod_{n=1}^5 (X-A_n)(X+A_n).
\ea
\ee
In our case, the branch points $A_n$ are positive real and we order them increasingly.
We define the $\mathcal A_1$ cycle to be the one that encircles $A_4$ and $A_5$ counterclockwise, the $\mathcal A_2$ cycle encircling $A_2$ and $A_3$ counterclockwise. We also define the $\mathcal B_1$ cycle encircling  $A_1$ and $A_4$, and the $\mathcal B_2$ cycle encircling $A_1$ and $A_3$. For a cycle $\mathcal C \in \{ {\mathcal A}_1, {\mathcal A}_2, {\mathcal B}_1,{\mathcal B}_2 \}$, we define
\be
\ba
	\Pi_{\mathcal C} &= \oint_{\mathcal C}  \left [  \log X \left ( -\frac{5}{2X}+\frac{X(5{X}^4+3 {X}^2(-\kappa^2+2\kappa_1)+\kappa_1^2}{2\sqrt{\sigma(X)}}  \right ) \right. \\
		& \qquad  \qquad  \qquad \qquad 
		\left. +\frac{\pi}{2} \frac{5{X}^2-3 {X}\kappa+\kappa_1}{({X}^2-X \kappa +\kappa_1) \sqrt{\sigma(X)}} \right ] \rd X, \\
	\Pi^{(\lambda)}_{\mathcal C} &= \oint_{\mathcal C} \frac{-2 \kappa X^3 \rd X}{\sqrt{\sigma(X)}}, \\
	\Pi^{(\lambda_1)}_{\mathcal C} &= \oint_{\mathcal C}  \frac{ 2X(X^2+\kappa_1)\rd X}{\sqrt{\sigma(X)}}.\\
\ea
\ee
\begin{figure}[t]
\begin{center}
\includegraphics[scale=1]{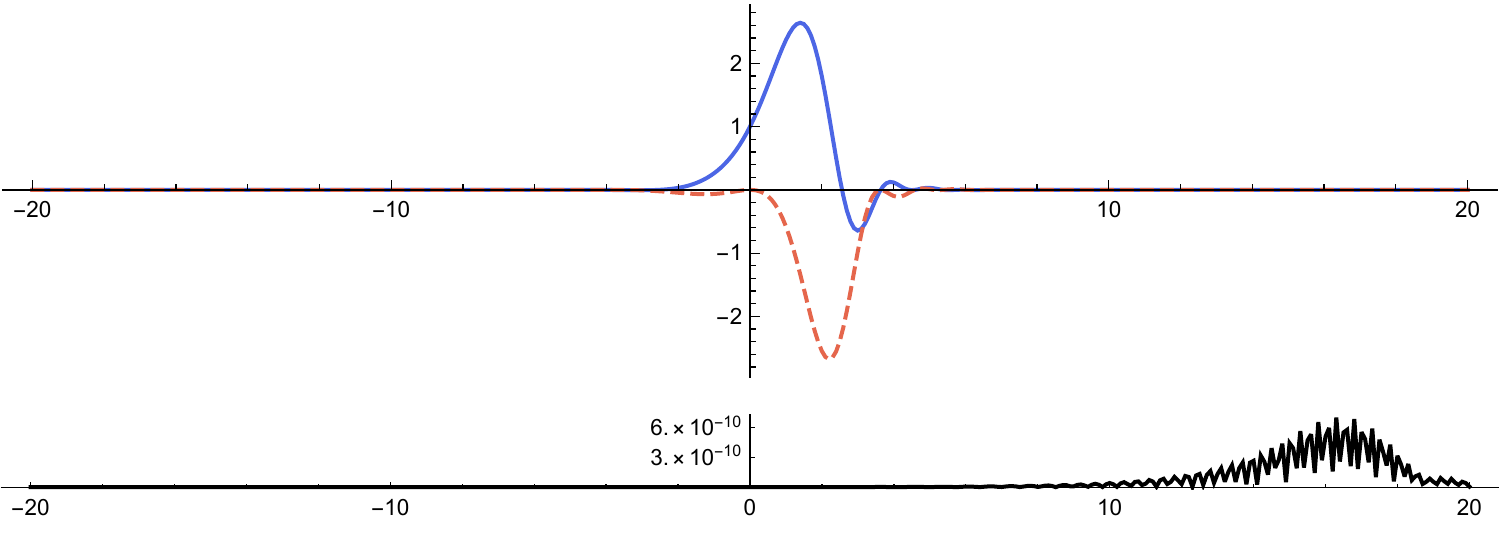}
\caption{Local $\mathbb C^3 / \mathbb Z_5$ at $\hbar=\pi$: exact $(0,0)$ eigenfunction (real part in blue and imaginary part in red) and the absolute difference with numerics coming from numerical diagonalization of a $250 \times 250$ matrix (rescaled to match the exact eigenfunctions at $x=0$). For this size of the matrix, the maximal difference is of the order $10^{-10}$. }
\label{c3z5plot00}
\end{center}
\end{figure}
\begin{figure}[t]
\begin{center}
\includegraphics[scale=1]{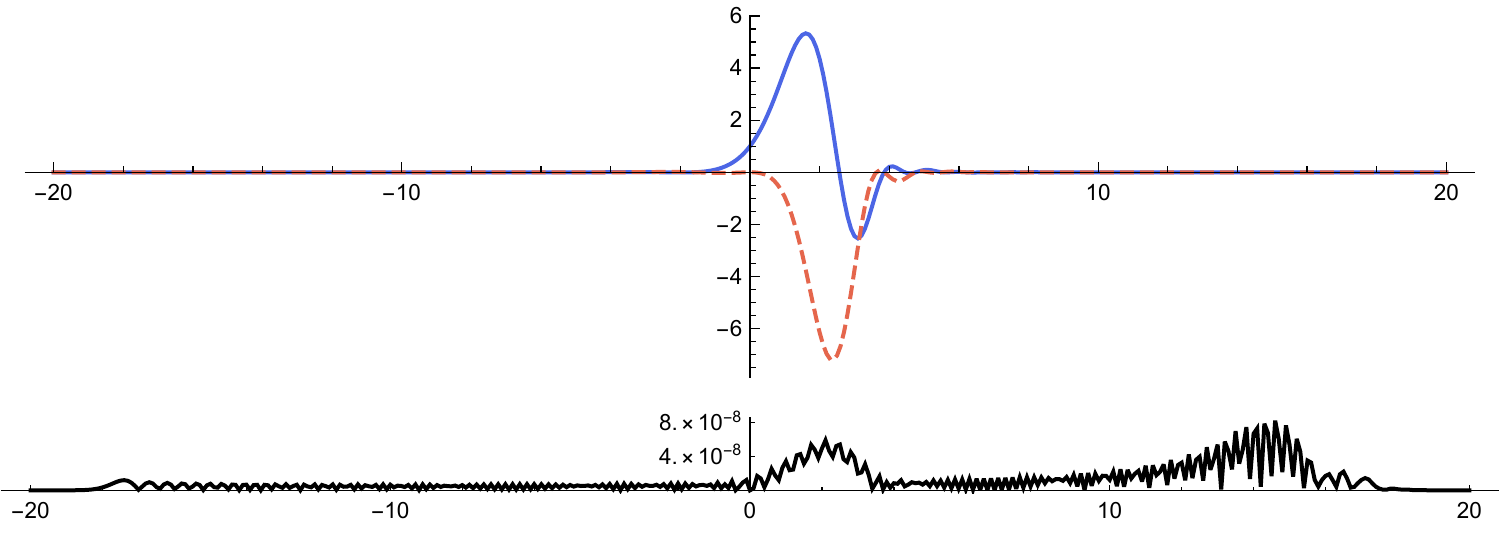}
\caption{Local $\mathbb C^3 / \mathbb Z_5$ at $\hbar=\pi$: same plot as in Fig. 6 for the exact $(0,1)$ eigenfunction. The maximal difference is of the order $10^{-8}$. }
\label{c3z5plot01}
\end{center}
\end{figure}
\begin{figure}[t]
\begin{center}
\includegraphics[scale=1]{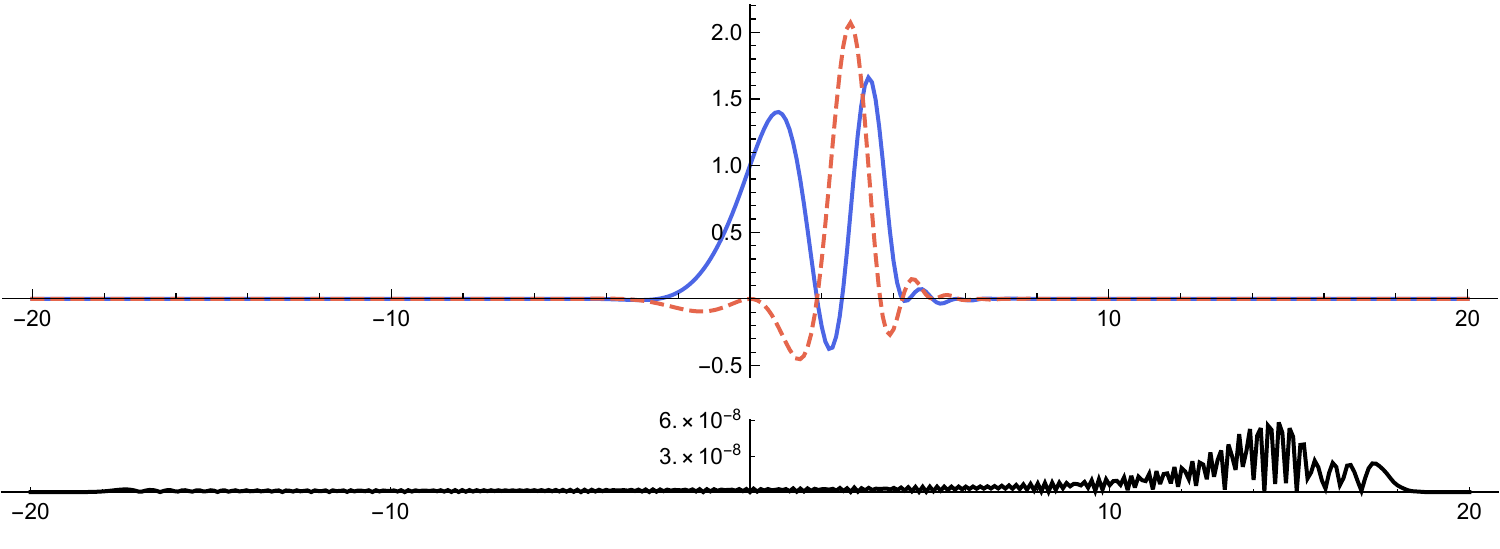}
\caption{Local $\mathbb C^3 / \mathbb Z_5$ at $\hbar=\pi$: same plot as in Fig. 6 for the exact $(1,0)$ eigenfunction. The maximal difference is of the order $10^{-8}$. }
\label{c3z5plot10}
\end{center}
\end{figure}
\noindent
Then, by monodromy invariance along all the periods, we have to solve
\be
	\label{c3z5mon1}
\ba
	\frac{\ri}{\pi} \left ( \Pi_{\mathcal A_i} + \lambda \Pi^{(\lambda)}_{\mathcal A_i}  + \lambda_1 \Pi^{(\lambda_1)}_{\mathcal A_i} \right )  = 2\pi \ri M_i ,\qquad i=1,2, \\
	\frac{\ri}{\pi} \left (  \Pi_{\mathcal B_i} + \lambda \Pi^{(\lambda)}_{\mathcal B_i}  + \lambda_1 \Pi^{(\lambda_1)}_{\mathcal B_i} \right )  = 2\pi \ri  N_i ,\qquad i=1,2. \\
\ea
\ee
From reality considerations, we find that $M_1=M_2=0$, so 
\be
	\begin{pmatrix*}[l]
		\lambda \\
		\lambda_1 
	\end{pmatrix*}
	=
	-\begin{pmatrix*}[l]
		\Pi^{(\lambda)}_{\mathcal A_1} &  \Pi^{(\lambda_1)}_{\mathcal A_1}  \\[0.1cm]
		\Pi^{(\lambda)}_{\mathcal A_2} &  \Pi^{(\lambda_1)}_{\mathcal A_2} 
	\end{pmatrix*}^{-1}
	\begin{pmatrix*}[l]
		\Pi_{\mathcal A_1}  \\
		\Pi_{\mathcal A_2} 
	\end{pmatrix*}.
\ee
Then, the two equations in the second line of (\ref{c3z5mon1}) give the quantization conditions for $\kappa$ and $\kappa_1$. Let us define $N_1=n_1+n_2+2$, and $N_2=n_2+1$. We find
\be
\ba
	(n_1,n_2)&=(0,0), \qquad \quad (\kappa,\kappa_1)_{0,0} = (\,\, -12.108260493297777783\ldots,
											    \,\,  14.559207122129454382\ldots),\\
	(n_1,n_2)&=(1,0), \qquad \quad (\kappa,\kappa_1)_{1,0} = (\,\, -25.786257255292478834\ldots, 
											    \,\,  19.332115669949562433\ldots),\\
	(n_1,n_2)&=(0,1), \qquad \quad (\kappa,\kappa_1)_{0,1} = (\,\, -17.514284419555278234\ldots, 
											     \,\, 40.988214882454531661\ldots),\\
	\ldots 
\ea
\ee
and so on. The case $(n_1,n_2)=(1,0)$ has been computed in \cite{fhm} by other means, and our result matches it. Some eigenfunctions and the comparison with numerical results can be seen in Fig. \ref{c3z5plot00},  \ref{c3z5plot01} and \ref{c3z5plot10}. We proceed in the following way: we fix the value of $\kappa_1$ and perform the numerical diagonalization with respect to $\kappa$, using hamiltonian truncation as usual. We find good agreement for both eigenvalues and eigenfunctions.

 
 
 \sectiono{Comparison with conjectured quantization condition}
\label{sect5}

As already mentioned in the introduction, there are conjectural expressions for the quantization conditions related to our family of difference equation. 
These quantization conditions were first presented in \cite{wzh} for some simple cases (where the underlying mirror curve has genus one)  and in \cite{hm,fhm} for more general cases. They are built using the refined topological string partition function in the so called Nekrasov-Shatashvili (NS) limit. 
This function is exact in $\hbar$, but is usually only known in the large-radius expansion, which corresponds to the large eigenvalue expansion. This expansion is a convergent series when $\hbar$ is real.
The exact quantization conditions found in the present work and the conjectured ones give the same spectrum in all the examples we tested: they both correctly reproduce the numerical results.
In this section, we provide numerical evidence that the two functions not only agree on the spectrum, but for any value of $\kappa$. Our exact expressions are in fact resummations of the conjectured quantization conditions in the rational case. 
For concreteness, we take the difference equation of local $\mathbb P^1 \times \mathbb P^1$.

The conjectured quantization condition for local $\mathbb P^1 \times \mathbb P^1$ has been written down in  \cite{wzh}, and it is the $N=2$ case of \cite{hm}. The instanton part of the refined free energy of local $\mathbb P^1 \times \mathbb P^1$ can be computed for example in the refined topological vertex formalism \cite{ikv}. In the NS limit, it has the form
\be
	F_{\rm NS}(T_1,T_2,\hbar) = -\sum_{w=1}^{\infty} \sum_{d_1,d_2 = 0}^{\infty} \, \sum_{j_L, j_R} N^{d_1,d_2}_{j_L, j_R} \frac{ \sin \frac{\hbar w}{2}(2 j_L+1) \sin \frac{\hbar w}{2}(2 j_R+1) }{2 w^2 \sin^3 \frac{\hbar w}{2} } \re^{- w d_1 T_1} \re^{- w d_2 T_2},
\ee
where the BPS invariants $N^{d_1,d_2}_{j_L, j_R} $ are integer numbers (determined by the geometry of the underlying toric CY threefold) which are counting BPS states of the M-theoretic lift of the topological string. In addition to the wrapping numbers $d_1$ and $d_2$, they depend on spin numbers $j_L$ and $j_R$ which are half-integers. The first BPS invariants for local $\mathbb P^1 \times \mathbb P^1$ can be found for example in \cite{ckm}. We also define the following derivative:
\be
	\frac{\partial F_{\rm NS}}{\partial T}(T_1,T_2,\hbar) = \frac{\partial F_{\rm NS}(T_1,T_2,\hbar) }{\partial T_1}+\frac{\partial F_{\rm NS}(T_1,T_2,\hbar) }{\partial T_2}.
\ee
The $T_i$ are the K\"ahler parameters of the geometry. They will be replaced by the ``quantum mirror map" $\hat T_i$ which is a deformation of the standard mirror map relating the K\"ahler parameters to the moduli and parameters of the mirror curve, here $\kappa$ and  $m$. The quantum mirror map for local $\mathbb P^1 \times \mathbb P^1$ has been constructed in \cite{acdkv}. Actually, it can be easily obtained from the resummed WKB eigenfunction computed previously, using the $g_k(\tilde X)$ in (\ref{gkfunctions}):
\be
\ba
	\hat T_2 &=  \log \kappa^2+2 \underset{{\tilde X=0}}{\rm Res} \left ( \frac{1}{\tilde X} \sum_{k \geq 1}g_k(\tilde X)\kappa^{-k} \right ), \\
		&= \log \kappa^2 -\frac{2m+1}{\kappa^2}-\frac{3m^2+2m(4+\re^{\ri \hbar}+\re^{-\ri \hbar})+3}{\kappa^4}+...  \\
	\hat T_1 &= \hat T_2 - \log(m).
\ea
\ee
We also need the classical volume of space phase, which can be computed using semiclassical methods (it is also given in \cite{wzh,hm}). In our case, we find
\be
	\frac{1}{2\pi \hbar} \log(\kappa^{2})^2-\frac{1}{2\pi \hbar}\log(\kappa^{2})\log(m)-\frac{\pi}{3\hbar}-\frac{\hbar}{12\pi}.
\ee
The quantum corrected volume is then conjectured to be given by the classical volume where $\kappa$ is replaced by the quantum mirror map, plus a contribution coming from the NS free-energy. The quantization condition is then conjecturally given by
\be
\ba
	n+\frac{1}{2}  &= \frac{1}{2\pi \hbar} \hat T_2^2-\frac{1}{2\pi \hbar}\hat T_2 \log(m)-\frac{\pi}{3\hbar}-\frac{\hbar}{12} \\
	& \qquad + \frac{\partial F_{\rm NS}}{\partial T}(\hat T_1,\hat T_2,\hbar)+\frac{\partial F_{\rm NS}}{\partial T}\left ( \frac{2\pi}{\hbar} \hat T_1, \frac{2\pi}{\hbar} \hat T_2, \frac{4\pi^2}{\hbar} \right ).
\ea
\ee
This expression is manifestly self dual under the duality transformation $\hbar/2\pi \leftrightarrow 2\pi/\hbar$ and \mbox{$\hat T_i \rightarrow \frac{2\pi}{\hbar} \hat T_i$}, $m \rightarrow m^{2\pi/\hbar}$.
In the rational case when $\hbar/2\pi \in \mathbb Q$, the two expressions involving $F_{\rm NS}$ have poles. 
But the singular parts exactly cancel and the overall result is finite. Note that during this manipulation, the $\hat T_i$ are assumed to be fixed (we do not vary their $\hbar$ dependance).
\begin{table}[t]
\begin{center}
\begin{tabular}{c  l}
order in $\kappa^{-1}$ & value for $\kappa=-33$ and $m=7$ \\[0.05cm]
\hline
 0 &  \underline{2. 1}26 503 418 773 313 1 \\[0.05cm]
 2  & \underline{2. 114} 410 838 199 331 8 \\[0.05cm]
 4  & \underline{2. 114 29}2 905 714 983 6 \\[0.05cm]
 6  & \underline{2. 114 290 9}69 742 037 0 \\[0.05cm]
 8  & \underline{2. 114 290 93}1 400 628 3 \\[0.05cm]
10 & \underline{2. 114 290 930 5}67 097 7 \\[0.05cm]
12 & \underline{2. 114 290 930 547} 962 6 \\[0.05cm]
14 & \underline{2. 114 290 930 547} 507 2 \\[0.05cm]
16 & \underline{2. 114 290 930 547 496} 5 \\[0.05cm]
\hline
exact value 
     & 2. 114 290 930 547 496 280 8...
\end{tabular}
\caption{The values of the RHS of the quantization condition (\ref{qcApprox}) for $m=7$ and off-shell value $\kappa=-33$. The expression is truncated after the indicated order in large $\kappa$. It converges to the value obtained by the evaluation of the exact expression (\ref{qcExact}) (the periods are evaluated numerically). \label{tab1}}
\end{center}
\end{table}
 After pole cancellation, the $\kappa$ expansions for $\hat T_i$ can be plugged in the expression, and we obtain a quantization condition for $\kappa$ as a large $\kappa$ series. 
 For example, for the case
 \be
 	\hbar =2\pi/3,
 \ee
  we find
 \be
	 \label{qcApprox}
 \ba
 	n+\frac{1}{2} &= \frac{3 \log \left(\frac{1}{\kappa ^2}\right) \left(\log \left(\frac{1}{\kappa
   ^2}\right)+\log (m)\right)}{4 \pi ^2}-\frac{5}{9} 
   +\frac{(m+1) \left(18 \log \left(\frac{1}{\kappa ^2}\right)+9 \log (m)+2 \sqrt{3} \pi \right)}{6 \pi ^2
   \kappa ^2} \\
   & \quad +\frac{3 (m+1)^2 \left(6 \log \left(\frac{1}{\kappa ^2}\right)+3 \log
   (m)+4\right)+\frac{2 \pi  (m (3 m+4)+3)}{\sqrt{3}}}{4 \pi ^2 \kappa
   ^4} +\ldots 
 \ea
 \ee
 This expression is found to converge for large enough $|\kappa|$, and we can use it to find the spectrum $\kappa_n$ for $n=0,1,2,\ldots$. 
 
  Let us recall the exact quantization condition in terms of period integrals, which we found in (\ref{quantp1xp1}). It can be written as
 \be
 	 \label{qcExact}
 \ba
 	n+\frac{1}{2} &= \frac{3}{4\pi^2} \left (  \Pi_{B}-\left (\frac{\Pi_A+m \log (m) \Pi_A^{(\lambda_m)}}{\Pi_{A}^{(\lambda)}} \right )\Pi^{(\lambda)}_{B}
	+m \log(m) \Pi_{B}^{(\lambda_m)} \right ) -\frac{1}{2}.
\ea
 \ee
 Both expressions (\ref{qcApprox}) and (\ref{qcExact}) give the same spectrum. Moreover, the righthand sides agree also when $\kappa$ does not satisfy the quantization condition, as can be seen in Table \ref{tab1}. The second expression is effectively the resummation of the first expression.
In this example, this can be also checked in the large $\kappa$ expansion by explicitly expanding the period integrals (\ref{periods2pio3}). 
Since in the large $\kappa$ limit the $A$ cycle shrinks to a small cycle around a pole, the large $\kappa$ expansion of the different $A$ period integrals can be computed using residues.
The $B$ period integrals can be obtained in the large $\kappa$ expansion using for example the Picard-Fuchs equations.  Let us define the following basic period integrals\footnote{These are essentially the classical periods of the curve.} for local $\mathbb P^1 \times \mathbb P^1$:
\be
\ba
	\label{basicper}
	\pi_{A,B}(\kappa,m) &= -\oint_{\mathcal A,\mathcal B} \log(X) \left (\frac{{X}^2-m}{X\sqrt{-4X^2+(m+X(X+\kappa))^2}}\right )\rd X \\
	&= \oint_{\mathcal A,\mathcal B} \log \left ( \frac{m+X^2+\kappa  X+\sqrt{-4X^2+(m+X(X+\kappa))^2}}{2 X} \right )\frac{\rd X}{X}.
   \ea
\ee
They satisfy the Picard-Fuchs equation given by
\be
\ba
	0&=\kappa^3 \pi_{A,B}'-(16(m-1)^2+8(m+1)\kappa^2-3\kappa^4) \pi_{A,B}'' \\
	& \qquad \qquad \qquad \qquad+\kappa((\kappa-2)^2-2m)((\kappa+2)^2-2m)\pi_{A,B}''',
\ea
\ee
where $'$ denotes the derivative w.r.t. the modulus $\kappa$.
We find
\be
\ba
 \pi_A(\kappa,m) &=\ri \pi  \log \left(\frac{1}{\kappa ^2}\right)+\frac{2 \ri \pi  (m+1)}{\kappa ^2}+\frac{3 \ri \pi  \left(m^2+4 m+1\right)}{\kappa ^4}+\mathcal O(\kappa^{-6}),\\
  \pi_B(\kappa,m) &=\log ^2\left(\frac{1}{\kappa ^2} \right)+\log \left(\frac{1}{\kappa ^2}\right) \log (m)-\frac{2 \pi ^2}{3}+\frac{2 (m+1) \left(2 \log \left(\frac{1}{\kappa ^2}\right)+\log (m)+2\right)}{\kappa ^2} \\
  	& \qquad
	+ \frac{6 \left(m^2+4 m+1\right) \log \left(\frac{1}{\kappa ^2}\right)+13 m^2+3 \left(m^2+4 m+1\right) \log
   (m)+40 m+13}{\kappa ^4}\\
   & \qquad +\mathcal O(\kappa^{-6}).
 \ea
\ee
The constant $-\frac{2 \pi ^2}{3}$ in the $B$ period can be fixed numerically for example.
Using these expansions, we can obtain large $\kappa$ series for (\ref{periods2pio3}) through the relations
\be
\ba
	\Pi_{A,B}^{(\lambda)} &=\frac{1}{9} \frac{\rd }{\rd \kappa} \pi_{A,B}(\kappa^3-3(m+1)\kappa,m^3), \\
	\Pi_{A,B}^{(\lambda_m)} &= \frac{1}{9} \frac{\rd }{\rd m} \pi_{A,B}(\kappa^3-3(m+1)\kappa,m^3).
\ea
\ee
The periods in the first line of (\ref{periods2pio3}) are made of two parts. The first part is easily expressed as
\be
	\oint_{{\mathcal A},{\mathcal B}} \left [ \log(X) \left ( -\frac{1}{X}+\frac{{X}^6-m^3}{X\sqrt{\sigma(X)}}\right )  \right ] \rd X =-\frac{1}{9}\pi_{A,B}(\kappa^3-3(m+1)\kappa,m^3).
\ee 
It turns out that the derivative of the second part of (\ref{periods2pio3}) containing the polynomials $p_i(X)$ can be obtained by applying a differential operator on $\pi_{A,B}$.
Indeed, we find that
\be
	\label{intp1op2kappa}
\ba
	&\frac{\rd}{\rd \kappa} \oint_{{\mathcal A}, {\mathcal B}}\frac{p_1(X)}{p_2(X)\sqrt{\sigma(X)}} \rd X = \\
	& \qquad  \qquad \qquad
	 \frac{4 m \left(3 \kappa ^4-9 \kappa ^2+2 m^3-3 \kappa ^2
   m^2+\kappa ^4 m-12 \kappa ^2 m-2\right)}{9 \kappa 
   \left(-\kappa ^2+m+1\right) \left(-\kappa ^2+3 m+3\right)} \\
   & \qquad \qquad \qquad \qquad \qquad \quad
   \times \frac{\rd}{\rd \kappa} \left ( \frac{1}{\kappa^2-m-1} \frac{\rd}{\rd \kappa}  \pi_{A,B}(\kappa^3-3(m+1)\kappa,m^3) \right ) \\
   &\qquad  \qquad \qquad
  + \frac{4}{9}m(m+1)  \frac{\rd}{\rd m} \left ( \frac{1}{\kappa^2-m-1} \frac{\rd}{\rd \kappa}  \pi_{A,B}(\kappa^3-3(m+1)\kappa,m^3) \right ).
\ea
\ee
The integration with respect to $\kappa$ can be performed by integrating the large $\kappa$ series for the r.h.s. term by term. The extra constants of integrations are found to be
\be
	c_A=0, \qquad \qquad c_B=-\frac{4\pi}{3\sqrt{3}},
\ee
for the $A$ and $B$ periods respectively. The $c_A$ can be found by performing a residue computation, whereas $c_B$ is found numerically.
Putting everything together, we can work out the large $\kappa$ expansion of (\ref{qcExact}). It precisely matches the expansion (\ref{qcApprox}) predicted by the conjecture of \cite{wzh}.

It is rather interesting to see that the complicated period integrals given by the l.h.s. of (\ref{intp1op2kappa}) can be given by a differential operator applied on the basic periods (\ref{basicper}) with the moduli and parameters substituted. This property is not clearly seen\footnote{at least by the author} from the algorithm of section \ref{sect3} producing the expressions for the l.h.s. of (\ref{intp1op2kappa}). However, it is tempting to suppose that this is a general feature, and therefore, that the exact quantization conditions for any $\hbar \in 2\pi \mathbb Q$ can be expressed in terms of some basic $\hbar$ independent periods (like (\ref{basicper}) for local $\mathbb P^1 \times \mathbb P^1$), together with the data given by a) the appropriate $\hbar$ dependent transformations of the moduli and parameters, b) finite differential operators involving only rational functions of the moduli and parameters, and c) an integration involving some integration constants.
This was checked for the simpler case $\hbar=\pi$ for local $\mathbb P^1 \times \mathbb P^1$.

 
 \sectiono{Conclusion}
 
 In this paper, we have built on several ideas and put them together in order to construct exact quantization conditions and eigenfunctions for a family of difference operators in the case where $\hbar/2\pi \in \mathbb Q$.
 These difference operators appear in the context of quantized mirror curves, and cluster integrable systems.
  The approach taken is constructive and not completely rigorous, but the results have passed all numerical tests, and agree with the existing results in the literature. These eigenfunctions, that we call ``fully on-shell", are a subset of the on-shell (or square integrable) eigenfunctions of the difference operator, and are relevant in the realm of integrable systems. In our construction, a central role is played by the Faddeev modular duality structure underlying the operators under considerations.
 
 Our discussion was restricted to the hyperelliptic cases, where the difference equation is of order 2. This was merely for technical reasons, and a priori, we do not expect conceptual difficulties in the treatment of higher order cases. Nevertheless, an explicit check is desirable. 
 
 In the context of the TS/ST correspondence, it appears that not only the so called ``fully on-shell" eigenfunctions play a role, but rather a more general set of functions, where neither $\kappa$ nor the other true moduli are restricted to a discrete set of values (see \cite{mz1,mz2}). It would be thrilling to have a first principle characterisation of those eigenfunctions too, since those eigenfunctions seem to be a cornerstone towards a non-perturbative definition of open topological string amplitudes on toric CY threefolds \cite{mz1,mz2}.
 

 \section*{Acknowledgements}
We would like to thank Santiago Codesido and Rinat Kashaev for stimulating discussions, and in particular Marcos Mari\~no for many helpful comments and careful reading of the draft.
This work is supported in part by the Fonds National Suisse, 
subsidies 200021-156995 and 200020-141329, and by the NCCR 51NF40-141869 ``The Mathematics of Physics" (SwissMAP).

\end{document}